\documentclass[onecolumn]{aastex7}
\usepackage{CJK}
\usepackage{bm}
\usepackage{caption}
\usepackage{subcaption}
\usepackage{appendix}
\usepackage[nolist,nohyperlinks]{acronym}

\begin{document}
	\begin{CJK*}{UTF8}{gbsn}
		
		\title{Abundance Pattern Fitting with Bayesian Inference: Constraining First Stars' Properties and Their Explosion Mechanism with Extremely Metal-poor Stars}
		
		\author[0009-0001-0604-072X]{Ruizheng Jiang}
		\affiliation{CAS Key Laboratory of Optical Astronomy, National Astronomical Observatories, Chinese Academy of Sciences, Beijing 100101, China
		}
		\affiliation{School of Astronomy and Space Science, University of Chinese Academy of Sciences, Beijing 100049, China
		}
		\email{jiangrz@bao.ac.cn}
		
		\author[0000-0002-0389-9264]{Haining Li}
		\affiliation{CAS Key Laboratory of Optical Astronomy, National Astronomical Observatories, Chinese Academy of Sciences, Beijing 100101, China
		}
		\email[show]{lhn@bao.ac.cn}
		
		\author[0000-0002-8980-945X]{Gang Zhao}
		\affiliation{CAS Key Laboratory of Optical Astronomy, National Astronomical Observatories, Chinese Academy of Sciences, Beijing 100101, China
		}
		\affiliation{School of Astronomy and Space Science, University of Chinese Academy of Sciences, Beijing 100049, China
		}
		\email[show]{gzhao@nao.cas.cn}
		
		\author[0000-0003-0663-3100]{Qianfan Xing}
		\affiliation{CAS Key Laboratory of Optical Astronomy, National Astronomical Observatories, Chinese Academy of Sciences, Beijing 100101, China
		}
		\email{qfxing@nao.cas.cn}
		
		\author[0000-0003-3646-9356]{Wenyu Xin}
		\affiliation{Department of Astronomy, Beijing Normal University, Beijing 100875, China}
		\affiliation{Institute for Frontiers in Astronomy and Astrophysics,
			Beijing Normal University, Beijing 102206, China}
		\email{xinwenyu16@mails.ucas.ac.cn}

		\begin{abstract}
			
			The abundance patterns of extremely metal-poor stars preserve a fossil record of the Universe's earliest chemical enrichment by the supernova explosions from the evolution of first generation of stars, also referred to as Population III (or Pop III).
			By applying Bayesian inference to the analysis of abundance patterns of these ancient stars, this study presents a systematic investigation into the properties and explosion mechanism of Pop III stars. 
			We apply NLTE corrections to enhance the reliability of abundance measurements, which significantly reduces the discrepancies in abundances between observations and theoretical yields for odd-Z elements, such as Na and Al. 
			Our Bayesian framework also enables the incorporation of explodability and effectively mitigates biases introduced by varying resolutions across different supernova model grids. 
			In addition to confirming a top-heavy ($\alpha=0.54$) initial mass function for massive Pop III stars, we derive a robust mass--energy relation ($E\propto M^2$) of the first supernovae. 
			These findings demonstrate that stellar abundance analysis provides a powerful and independent approach for probing early supernova physics and the fundamental nature of the first stars.
			
		\end{abstract}
		
		\keywords{\uat{Stellar abundances}{1577} --- \uat{Population III stars}{1285} --- \uat{Stellar mass functions}{1612} --- \uat{Core-collapse supernovae}{304}}
		
		\section{Introduction} 
		\label{sec:intro}
		
		In near-field cosmology, low-metallicity stars are treated as local equivalents to the high-redshift Universe to understand the earliest metal enrichment \citep{frebel_near-field_2015}. 
		Metal-poor stars with metallicity $\mathrm{[Fe/H]}<-2$ and $\mathrm{[Fe/H]}<-3$ are referred to as \ac{VMP} and \ac{EMP} stars \citep{beers_discovery_2005, bonifacio_most_2025}.
		After excluding chemical contamination from both possible mass transfer in a binary system \citep[e.g.][]{bisterzo_s_2010, bisterzo_s-process_2011} or internal processes occurring during specific evolutionary phases \citep{placco_carbon-enhanced_2014}, the observed abundances of a star can be regarded as a reliable record of the chemical composition of its birth environment. 
		Given that massive stars have short lifespans and contribute to the early chemical enrichment through their explosive deaths in the primordial Universe, the abundance patterns preserved in \ac{EMP} stars are commonly used to investigate the properties of the first-generation massive stars and their associated supernova explosions \citep[e.g.][]{placco_metal-poor_2015, placco_observational_2016, hartwig_descendants_2018, vanni_characterizing_2023, koutsouridou_true_2024, ji_spectacular_2024}. 
		By comparing the chemical abundances of \ac{EMP} stars and the theoretical yields, the enriching event to the birth environments where these ancient stars formed could be inferred. 
		
		High-resolution spectroscopy with resolving power at $R>20000$ enables precise measurements of the chemical abundance patterns of the \ac{EMP} stars.
		Extensive observations have been conducted over the past two decades, which provides profound insights into the abundance patterns and chemical origins of \ac{VMP} and \ac{EMP} stars, including the samples of First Stars \citep[hereafter FS]{cayrel_first_2004, bonifacio_first_2009}, the Most Metal-poor Stars \citep[hereafter MMP]{norris_most_2013, yong_most_2013}, and Four-hundred Very Metal-poor Stars Studied with LAMOST and Subaru \citep[hereafter VMP 400]{aoki_four-hundred_2022, li_four-hundred_2022}.
		Abundance analysis has witnessed great progress in atmospheric model and synthesis theory of \ac{NLTE} effect for FGK-type stars \citep{lind_three-dimensional_2024}. Although the application of \ac{NLTE} radiative transfer in three-dimensional modeling is still numerically expensive, one-dimensional \ac{NLTE} grids \citep[e.g.][]{lind_non-lte_2022} is now feasible to abundance analysis and helps achieve more accurate measurement.
		To derive the properties of their progenitors, it is a common practice to match the observed abundances against theoretical supernova yields of zero-metallicity massive stars.
		Stellar evolution up to the pre-supernova phase has been extensively investigated, supported by a variety of well-structured programs with sophisticated nuclear reaction networks, such as KEPLER \citep{weaver_presupernova_1978, woosley_evolution_2002}, FRANEC \citep{limongi_evolution_2003, limongi_nucleosynthesis_2006} and MESA \citep{paxton_modules_2011}. 
		However, the determination of explosion mechanism for these massive stars remains a key challenge. 
		Although still debated, predictions based on the neutrino heating mechanism roughly align with current observations of compact remnant masses, kicks, and spins \citep{janka_explosion_2012}. 
		Recent advances in three-dimensional simulations offer promising avenues for achieving more physically consistent \ac{CCSN} models \citep{burrows_core-collapse_2021}. 
		However, such simulations are computationally intensive and not yet feasible for grid-based nucleosynthesis yield calculations. 
		Consequently, current yield models still rely on one-dimensional stellar evolution and arbitrary triggered explosions \citep{umeda_nucleosynthesis_2002, heger_nucleosynthesis_2010, limongi_presupernova_2012}. 
		
		The fate of Pop III massive stars has been long discussed.
		A star with an initial mass in the range of $10\,M_\odot$ to $100\,M_\odot$ ends its life with a gravitation-induced core collapse.
		Whether it explodes as a supernova or falls into a black hole (also named as failed supernova) depends on its initial mass and explosion energy.
		Other factors, including stellar rotation and magnetic field, are not discussed in this work, although they are also responsible for the outcome \citep{burrows_core-collapse_2021}. In general, a core-collapse explosion would occur for $9\,M_\odot<M<40_\odot$, while a direct black hole formation without supernova explosion for $M>40\,\odot$ \citep{klessen_first_2023}. However, the non-monotonic behavior of supernova explodability are repeatedly reported in the last decade \citep[e.g.][]{oconnor_black_2011, ertl_two-parameter_2016}. Considering that the interstellar medium could only be enriched by successful supernova explosions, this non-monotonicity would inevitably challenge the derivation of \ac{IMF} from the abundance patterns of \ac{EMP} stars. Additionally, recent researches have also highlighted the significance of multi-enrichment scenario where multiple supernovae could contribute to the pristine gas in the early Universe \citep[e.g.][]{hartwig_machine_2023}. A considerable fraction of \ac{PISN} with $140\,M_\odot<M<260\,M_\odot$ could be also predicted \citep{jiang_modified_2024} under this scenario. Given that the multiplicity of a certain star still remains inconclusive at present, this work only considers the enrichment from single supernova. 
		
		In this work, we only select \ac{EMP} star observations from homogeneous samples of  FS, MMP and VMP 400 (Section~\ref{sec:data.selection_criteria}). \ac{NLTE} effect is also analyzed with due consideration of systematic uncertainties from both atmospheric parameters and atmospheric models (Section~\ref{sec:data.nlte}). 
		We refined the methodology of the interpretation of observed abundance patterns with nucleosynthesis yields by incorporating Bayesian inference (Section~\ref{sec:method}), which provides a robust framework for deriving the progenitors of \ac{EMP} stars with consideration of both observational effects and grid resolutions of supernova models. A specially defined parameter, \textit{sensitivity}, is introduced for a quantitative assessment of the influence of each element on progenitor derivation (Section~\ref{sec:result.element}). This framework also contributes to reconcile the differences in resulting \ac{IMF} between \citet{ishigaki_initial_2018} and \citet{jiang_modified_2024} (Section~\ref{sec:result.imf_bias}). 
		Moreover, a well-constrained two-dimensional distribution of initial mass could be naturally obtained under this framework, thereby providing observational constraints on the mass--energy relation and the explodability of metal-free \ac{CCSN} (Section~\ref{sec:discuss.constraint_snmodel}). In addition, we apply the derived mass--energy relation on the updated explodability-modified IMF fitting (Section~\ref{sec:discuss.imf_edf_fit}). 
		
		\section{Data} 
		\label{sec:data}
		
		\subsection{Target Selection \& Element of Interest}
		\label{sec:data.selection_criteria}
		
		Due to the relatively limited number of existing high-resolution spectra, it usually requires compiling metal-poor stars from multiple researches or throughout databases, such as the SAGA \citep{suda_stellar_2008} and JINA \citep{abohalima_jinabasedatabase_2018}, to obtain a sample with a sufficient number of \ac{EMP} stars \citep[e.g.][]{fraser_mass_2017, ishigaki_initial_2018}.
		However, variations in measurement methods employed in different studies inevitably introduce discrepancies in the reported elemental abundances, thereby increasing the systematic uncertainty in the derivation of progenitor properties.
		In order to reduce systematic uncertainties as much as possible, we restrict our analysis to homogeneous \ac{EMP} samples with considerable numbers of stars, including FS, MMP and VMP 400.
		As a consequence of this strict sample selection, the number of targeted stars remains limited, which could be only addressed through the continued accumulation of high-quality, high-resolution, homogeneous spectroscopic observations in the future. 
		
		In addition to sample selection, we also impose extra criteria on the elemental species to be included in our analysis.
		Since metal-free supernovae are expected to produce only light elements up to zinc \citep{umeda_nucleosynthesis_2002, heger_nucleosynthesis_2010, limongi_presupernova_2012}, our chemical analysis is restricted to the light elements before zinc.
		Furthermore, we only include the species whose abundances are more frequently measured in high-resolution spectroscopic observations.
		These limitations lead to only eleven elements are included in our analysis, ranging from carbon, $\alpha$-elements (Mg, Si, Ca and Ti), iron-group elements (Mn, Fe, Co and Ni) and other odd-Z elements (Na and Al).
		Despite Sc and Cr are also easily detectable, they are disregarded due to the existence of large discrepancies between observational abundances and theoretical predictions \citep{limongi_presupernova_2012, kobayashi_origin_2020}.
		Therefore, the target stars are required to contain the measurements of the abundances of all the included eleven elements, for the analysis of individual effects of each element on the progenitor derivation. 
		
		Although the atmospheric abundance pattern records the chemical enrichment of interstellar medium at birth, it could be also altered by extrinsic contamination, including mass transfer from a companion star. For the purpose of avoiding metal contamination from mass transfer across a binary system, \ac{CEMP}-s (with $\mathrm{[Ba/Fe]}>+1.0$ and $\mathrm{[Ba/Eu]}>+0.5$) and \ac{CEMP}-r/s (with $0.0\leq\mathrm{[Ba/Fe]}\leq+0.5$) stars are discarded \citep{placco_carbon-enhanced_2014}. 
		Considering that Ba or Eu abundances are unavailable for some stars, we also examine astrometric binarity as a proxy for identifying non-single stars. 
		Specifically, we employed the \ac{RUWE} from Gaia Data Release 3 \citep{gaia_collaboration_gaia_2016, gaia_collaboration_gaia_2023}, and all sources in our sample satisfy $\mathrm{\ac{RUWE}} < 1.4$.
		For the VMP 400 sample, we further exclude seven double-lined spectroscopic binaries based on the binary analysis of \citet{aoki_four-hundred_2022}.

\begin{figure*}
			\centering
			\includegraphics[width=1\linewidth]{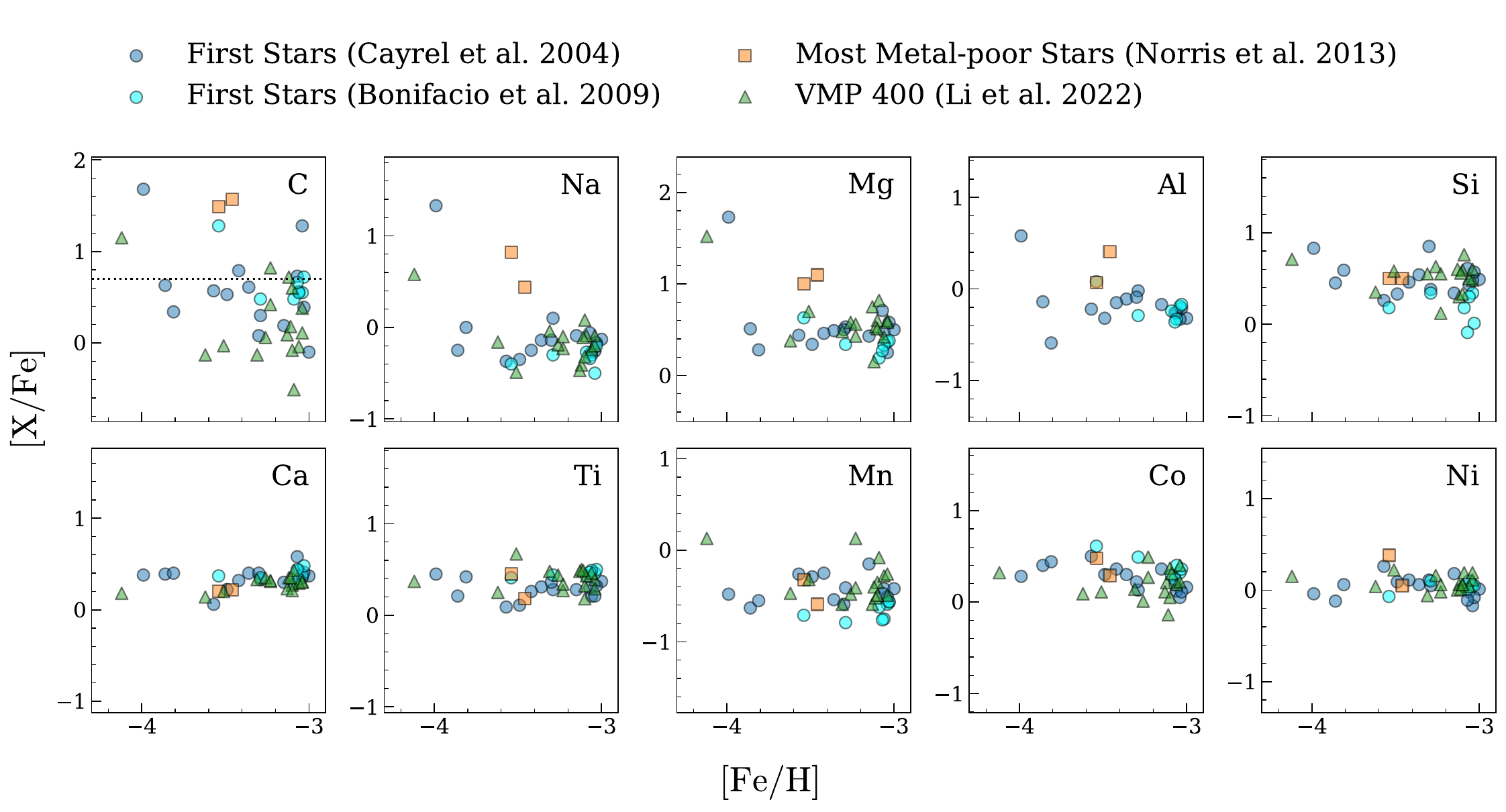}
			\caption{
				The abundances of EMP stars collected from FS, MMP and VMP 400 samples, which are represented as blue circles, orange squares and green triangles respectively.
				For FS samples, stars cataloged in different publications are distinguished by different brightness.
				The carbon abundance is corrected according to stellar evolution, while the sodium, magnesium and aluminum abundances are corrected for 1D \ac{NLTE} effect. 
				The typical observational uncertainties is on the order of $0.1\,\mathrm{dex}$. 
				The dotted horizontal line in the carbon panel represents the division of \ac{CEMP} of $\mathrm{[C/Fe]>0.7}$.
			}
			\label{fig:abudist}
		\end{figure*}

		For the FS, MMP and VMP 400 samples, the numbers of the selected \ac{EMP} stars based on the criteria above are limited to 22, 2 and 16 respectively, amounting to 40 out of 484, after \ac{NLTE} correction in Section~\ref{sec:data.nlte}. It should be noted that Al abundance is not an essential criteria sample selection. This is due to the fact that the detection of Al is hindered by the wavelength coverage of VMP 400 sample. The exclusion of Al abundance does not seriously affect the progenitor derivation after \ac{NLTE} correction (see Section~\ref{sec:result.element.nondetection}). The selected stars from these three homogeneous samples are cataloged in Table~\ref{tab:atmospheric_astrometric_parameter}, including the Gaia astrometric parameters and the \ac{LTE} atmospheric parameters, i.e. effective temperature $T_\mathrm{eff}$, surface gravity $\log g$, microturbulence $\xi_t$ and metallicity $\mathrm{[Fe/H]}$. The abundance distributions of the chosen elements of the selected stars from different samples are also demonstrated in Figure~\ref{fig:abudist}. It should be noted that the sodium, magnesium and aluminum abundances are corrected for 1D \ac{NLTE} effect (see Section~\ref{sec:data.nlte}). Additionally, the carbon abundance is also correct according to stellar evolution (see Section~\ref{sec:data.carbon_corr}).
		
		\startlongtable
		\begin{deluxetable*}{lcccccccccc}
			\label{tab:atmospheric_astrometric_parameter}
			\tablecaption{Basic Information, Astrometric Parameters and Atmospheric Parameters of the Selected Stars}
			\tablehead{
				\multicolumn{2}{c}{Observational Info.} & &\multicolumn{3}{c}{Astrometric Parameters (Gaia DR3)} & &\multicolumn{4}{c}{Atmospheric Parameters (\ac{LTE})} \\
				\cline{1-2} \cline{4-6} \cline{8-11}
				\colhead{Object name} &\colhead{Source} & &RA (J2000) &DEC (J2000) &RUWE & &\colhead{$T_\mathrm{eff}$} &\colhead{$\log g$} &\colhead{$\xi_t$} &$\mathrm{[Fe/H]}$ \\
				& & &h:m:s &d:m:s & & &K &[cgs] &$\mathrm{km\,s}^{-1}$ & 
			}
			\startdata
			BD-18 5550       &FS & &$19\ 58\ 49.74$ &$-18\ 12\ 11.14$ &0.993 & &4750 &1.4 &1.8 &$-3.06$ \\
			BS 16477-003 &FS & &$14\ 32\ 56.92$ &$+06\ 46\ 06.97$ &1.031 & &4900 &1.7 &1.8 &$-3.36$ \\
			CS 22172-002 &FS & &$03\ 14\ 20.85$ &$-10\ 35\ 11.28$ &1.191 & &4800 &1.3 &2.2 &$-3.86$\\
			CS 22186-025 &FS & &$04\ 24\ 32.80$ &$-37\ 09\ 02.52$ &1.000 & &4900 &1.5 &2.0 &$-3.00$\\
			\enddata
			\tablecomments{Table~\ref{tab:atmospheric_astrometric_parameter} is published in its entirety in the machine-readable format. A portion is shown here for guidance regarding its form and content.}
		\end{deluxetable*}
		
		\subsection{NLTE effect} 
		\label{sec:data.nlte}
		
		The \ac{NLTE} effect poses significant roles in abundance determinations. Tremendous efforts have been devoted to \ac{NLTE} calculation in multiple investigations for different elements, such as odd-Z elements Na \citep[e.g.][]{lind_non-lte_2011, alexeeva_non-lte_2014}and Al \citep[hereafter L22]{lind_non-lte_2022}, $\alpha$-elements O \citep[e.g.][]{sitnova_influence_2013, amarsi_galactic_2015}, Mg \citep[e.g.][]{osorio_mg_2015, osorio_mg_2016}, iron-peak elements Mn \citep[e.g.][]{bergemann_nlte_2008} and Fe \citep[e.g.][]{bergemann_non-lte_2012, lind_non-lte_2012}, and heavy elements Sr, Ba \citep[e.g.][]{mashonkina_even--odd_2019} and Eu \citep[e.g.][]{mashonkina_heavy_2001}. To ensure consistent \ac{NLTE} calculations, we adopt the most recent framework from \citet{lind_non-lte_2022}, which provides unified treatment for Na, Mg and Al, featuring extensive wavelength coverage from $2000\,$\r{A} to $3\,\mathrm{\mu m}$. 
		This \ac{NLTE} correction features fast computation of abundance corrections by interpolating the grids of observational atmospheric parameters and \ac{EW}s. Since it is carried out after the spectroscopic analysis is completed, a dependence on pre-determined atmospheric parameters would be inevitably introduced. The discrepancy in abundance correction could also arise due to different atomic data and atmospheric models. Therefore we analyze the systematic uncertainties introduced by atmospheric parameters and discrepancy across different \ac{NLTE} models. Our analysis of the \ac{NLTE} effect on abundances in this section includes the full FS, MMP, and VMP 400 samples to maximize the number of analyzed stars. The selection criterion of complete abundance measurements of the required elements are therefore not applied to the extended sample, as it will not affect the analysis of \ac{NLTE} effect in this section. 
		
		\subsubsection{Systematic Uncertainties}
		\label{sec:data.nlte.sysuncert}
		
		As previously stated, systematic uncertainties in abundance measurements would arise from both the atmospheric parameter determination and the choice of atmospheric models. In this section of systematic uncertainties, we aim to quantify and present the individual impact of each factor on the derived abundances.
		
		\paragraph{Atmospheric Parameters}
		\label{sec:data.nlte.sysuncert.atmparam}
		
		In the following, we focus on how the determination of atmospheric parameters affects \ac{NLTE} corrections. Among the required parameters, the effective temperature ($T_\mathrm{eff}$) and surface gravity ($\log g$) are particularly important for classifying the target stars. In this work, we adopt atmospheric parameters measured in previous studies. This also helps maintain a reliable correlation between atmospheric parameters and \ac{EW}s.
		
		The post-analysis \ac{NLTE} correction relies on atmospheric parameters pre-determined in the literature. In some cases, the stellar parameters are derived using spectroscopic methods under the \ac{LTE} assumption. It is recommended that atmospheric parameters be re-estimated from spectra under \ac{NLTE} assumptions to take consideration of the \ac{NLTE} effect on Fe (or Ti). Most of the stars analyzed in this study are, however, drawn from the MMP and VMP 400 datasets, where the atmospheric parameters are determined through methods independent of the \ac{LTE} assumption. Therefore, the \ac{NLTE} effect on atmospheric parameter determination is considered minor in this work.
		The differences in determination of these parameters however introduces systematic offsets across samples:
		\begin{itemize}
			\item \textbf{FS}: $T_\mathrm{eff}$ is estimated from the observed color indices, while $\log g$ is determined by balancing the Fe I and Fe II, Ti I and Ti II lines;
			\item \textbf{MMP}: $T_\mathrm{eff}$ and $\log g$ are both determined by spectrophotometry and hydrogen line profiles, while $T_\mathrm{eff}$ also combines H$\delta$ line indices;
			\item \textbf{VMP 400}: $T_\mathrm{eff}$ relies on color indices, with $\log g$ derived from parallax.
		\end{itemize}
		
		Thanks to different measurement methods adopted in \citet{norris_most_2013}, we could use the determination of $T_\mathrm{eff}$ and $\log g$ from both spectrophotometric flux and hydrogen line profiles. 
		The analysis of the influence of atmospheric parameters on abundance determination are detailed in Appendix~\ref{appx:atmparam_influence}. The corresponding results are summarized in Table~\ref{tab:compare_atm_impact}.
		Our analysis supports a more significant role of $T_\mathrm{eff}$ in abundance determination than $\log g$. 
		This \ac{lower sensitivity} to surface gravity measurement is also reported for the dwarf and subgiant subsample in \citet{yong_most_2013}, which share identical $T_\mathrm{eff}$ values but differing $\log g$ determinations. 
		The detailed discussion \ac{is} presented in Appendix~\ref{appx:atmparam_influence} together with the determined abundances and offset in atmospheric parameters listed in Table~\ref{tab:compare_atm_impact}.
		
		The difference in the determination of atmospheric parameters in general leads to an offset $\sim0.1\,\mathrm{dex}$ in measured abundances, which is relatively small compared to the \ac{NLTE} correction. 
		In addition, we can also roughly estimate the impact of atmospheric parameters on progenitor derivation by combining this abundance offset of $0.1\,\mathrm{dex}$ and the sensitivity of each element, as shown in Section~\ref{sec:result.element.sensitivity}. 
		Overall, the deviation in progenitor mass due to the atmospheric parameters approximates to $2\,M_\odot$. 
		It should be noted that, a systematic bias might also exist in $\xi_t$ \citep{roederer_search_2014}, while we only focus on the effects of $T_\mathrm{eff}$ and $\log g$ in this work.
		
		\paragraph{Model atmospheres}
		\label{sec:data.nlte.sysuncert.cog}
		
\begin{figure*}
			\centering
			\includegraphics[width=1.\linewidth]{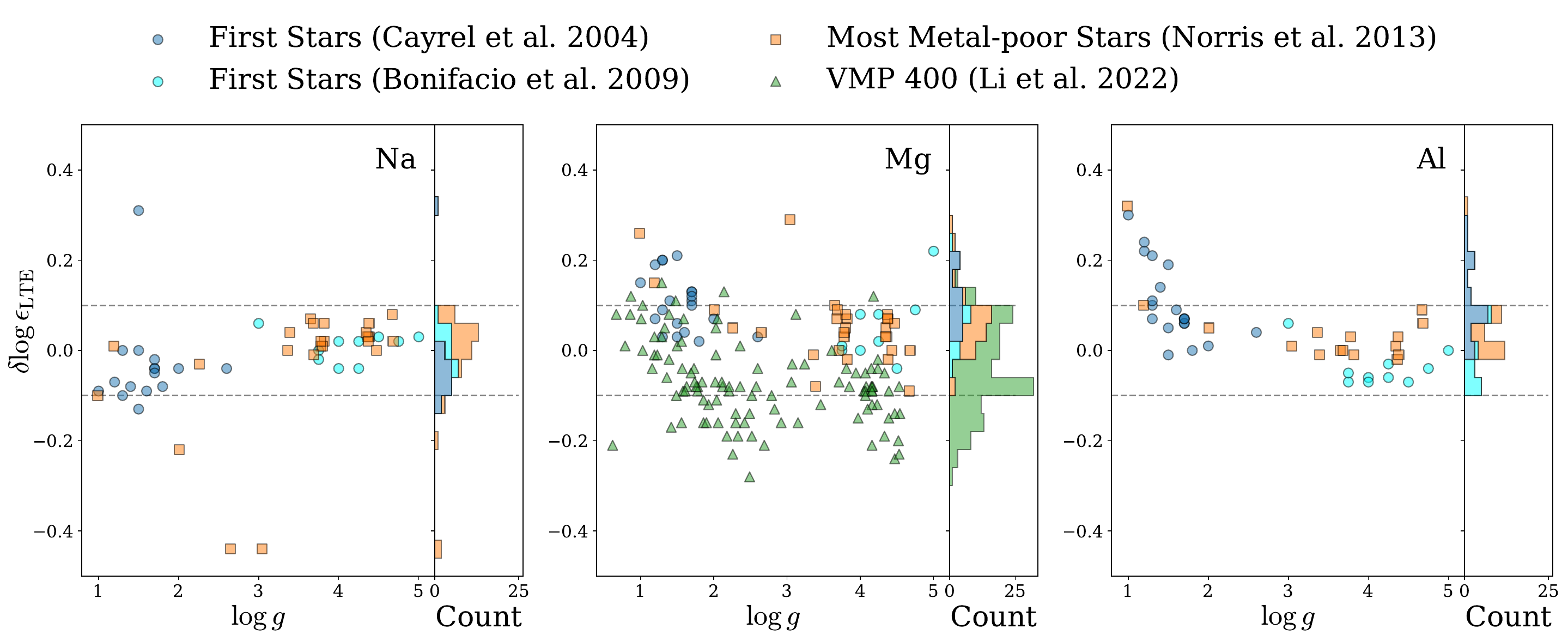}
			\caption{
				Deviation of the measured abundances $\delta{\log\epsilon}$, which is defined as Equation~\ref{eqn:delta_logeps}, as a function of surface gravity $\log g$ and their corresponding distributions.
				Target stars sourced from different samples are coded as the same in Figure~\ref{fig:abudist}.
				Abundance deviation $\delta{\log\epsilon}$ caused by different atmospheric models mostly falls within the typical observational uncertainty of $0.1\,\mathrm{dex}$, represented as th horizontal dashed lines. 
				However, the most significant deviations occur at the low surface gravity with $\log g<2$.
			}
			\label{fig:diff_cog}
		\end{figure*}

		The samples used in this study span nearly two decades of observations, during which the techniques of elemental abundance determination have undergone substantial revisions and updates. 
		Considerable effort has been made to shift toward more realistic frameworks based on \ac{NLTE} assumptions and three-dimensional hydrodynamical modeling \citep[see][]{lind_three-dimensional_2024}.
		Hence the difference in the adopted atmospheric models could also lead to systematic bias in abundance determination.
		The models and codes employed by the original papers are listed below for comparison:
		\begin{itemize}
			\item \textbf{FS}: OSMARCS model atmospheres \citep{gustafsson_grid_1975,gustafsson_grid_2008} with the spectral line analysis program TURBOSPECTRUM \citep{alvarez_near-infrared_1998};
			\item \textbf{MMP}: \citet{castelli_new_2003} model atmospheres with the MOOG spectrum synthesis program \citep{sneden_nitrogen_1973};
			\item \textbf{VMP 400}: \citet{castelli_new_2003} model atmospheres with the MOOG spectrum synthesis program \citep{sneden_nitrogen_1973}.
		\end{itemize}
		To quantitatively evaluate influence of differing atmospheric models, we calculate a new set of \ac{LTE} abundances for the target stars using the model grids provided by L22.
		We adopt the atmospheric parameters listed in Table~\ref{tab:atmospheric_astrometric_parameter} and take the \ac{EW}s from original literature for the consistency of the input parameters.
		The differences in measured abundances are defined as,
		\begin{equation}
			\label{eqn:delta_logeps}
			\delta{\log\epsilon_\mathrm{LTE}}=\log\epsilon_\mathrm{LTE}-\log\epsilon_\mathrm{literature}.
		\end{equation}
		As shown in Figure~\ref{fig:diff_cog}, it could be easily concluded that the difference in abundances $\delta{\log\epsilon}$ \ac{is} mostly beneath the typical observational uncertainty of $0.1\,\mathrm{dex}$. 
		However, an obvious elevation in the abundance deviation for Al exists at low $\log g$ end, extending to around $+0.4\,\mathrm{dex}$ at $\log g \simeq 0.5$. 
		This phenomenon could lead to a systematic bias in \ac{LTE} abundance measurement especially for the \citet{cayrel_first_2004} giant sample, which is represented with dark-blue circles in Figure~\ref{fig:diff_cog}. 
		
		\paragraph{Summary}
		The systematic bias could be attributed to the differences in both the determination of atmospheric parameter and the adoption of atmospheric model.
		In brief, it is necessary to ensure a homogeneous abundance analysis process under the same framework for maintaining consistency in the measured abundances in \ac{EMP} sample.
		Although decades of accumulation of high-spectroscopic observations have provided a large amount of abundance data for metal-poor stars, the homogeneous measurement of these high-quality spectra remains an important task, particularly when it comes to analyzing their chemical environment at birth.
		
		\subsubsection{Na, Mg \& Al}
		\label{sec:data.nlte.namgal}
		
		Na and Al abundances serve as critical diagnostics for distinguishing the chemical origins of stars, since their excess production needs surplus neutrons \citep{nomoto_nucleosynthesis_2013}.
		However, Na abundance would be overestimated by more than $+0.5\,\mathrm{dex}$ in \ac{LTE}-based analyses \citep[e.g.][]{lind_non-lte_2011, mashonkina_formation_2021}.
		Similarly, large discrepancies reaching $+0.9\,\mathrm{dex}$ are also observed in Al I and Al II abundances \citep{roederer_detection_2021}.
		These investigations together suggest that both Na and Al suffer from pronounced \ac{NLTE} effect.
		In addition, we also consider NLTE correction on Mg in this work, since it would also vary significantly between lines in giants \citep{osorio_mg_2015}.
		Different suites of \ac{NLTE} \ac{corrections} with publicly available online database based on distinct atmospheric models are developed, such as \citet{kovalev_mpia-nlte_2018} and \citet{mashonkina_1d_2023}.
		In order to prevent possible systematic correction between different atmospheric models, we utilize the updated L22 \ac{NLTE} correction based on the same framework. 
		
		The \ac{NLTE} abundances for Na, Mg and Al are generated by linearly interpolating the L22 correction grids of abundance and $\log\ac{EW}$.
		The detailed method and result of \ac{NLTE} correction and corresponding uncertainty are presented in Appendix~\ref{appx:nlte_corr}.
		Among the elements examined, both Na and Al exhibit substantial \ac{NLTE} effect, although the magnitude and behavior of this correction differ for different elements.
		For Na, the \ac{NLTE} correction becomes increasingly negative at higher \ac{LTE} abundance values.
		Notably, even though the Na abundances in the VMP 400 sample have already been corrected for \ac{NLTE} effects, a mean residual offset of approximately $-0.2\,\mathrm{dex}$ is still observed when compared with values derived using the updated \citet{lind_non-lte_2022} grids.
		In contrast, aluminum abundances exhibit a systematic positive correction, with a typical value around $+0.4\,\mathrm{dex}$, and a weaker correlation between the \ac{LTE} and \ac{NLTE} values than that seen for Na.
		For Mg, the \ac{NLTE} corrections are less pronounced but remain non-negligible.
		A correction ranging from approximately $+0.0$ to $+0.3\,\mathrm{dex}$ is also found, indicating that \ac{NLTE} effects must also be accounted for when precise Mg abundances are required in the study of stellar chemical abundance. 
		
		\subsubsection{Fe \& Ti}
		\label{sec:data.nlte.fe_ti}
		
		In addition to the aforementioned elements, Fe is a fundamental element, serving not only as the metallicity tracer of cosmic chemical evolution, but also as a critical input parameter of spectroscopic models for determinations of other elements.
		However, a significant variation in this metallicity indicator is observed across spectroscopic surveys: the FS sample adopts the mean value of [Fe I/H] and [Fe II/H], whereas both MMP and VMP 400 samples utilize [Fe I/H].
		Major \ac{EMP} compilations such as SAGA database preferentially adopt [Fe I/H] as the default metallicity indicator.
		This, to some \ac{extent, ensures} homogeneity across heterogeneous samples.
		
		Conventional stellar parameter analysis under LTE assumptions necessitates the demonstration of consistency between Fe I and Fe II abundances \citep{cayrel_first_2004}. However, current studies \citep{lind_non-lte_2011, bergemann_non-lte_2012} have revealed that Fe I abundances are significantly influenced by the \ac{NLTE} effects, rendering their \ac{LTE}-derived values potentially biased. 
		In contrast, Fe II abundances demonstrate comparatively minor \ac{NLTE} deviations \citep{bergemann_non-lte_2012}.
		
		We therefore utilize an alternative strategy to approaching the \ac{NLTE} effect by adopting [Fe II/H] as the metallicity, given that the \ac{NLTE} correction for Fe is not included in the L22 grids.
		However, this strategy is subject to observational limitations.
		The scarcity of detectable Fe II lines (particularly in metal-poor stars) has a substantial impact on the statistical reliability.
		In our homogeneous sample of $40$ selected stars, we compared the Fe I and Fe II abundances, finding that most Fe II abundances are consistent with those derived from Fe I.
		Consequently, we adopt [Fe I/H] as the iron abundance in the absence of Fe II abundances.
		Moreover, the same strategy is employed for the Ti abundance adoption.
		
		\subsection{Carbon Correction}
		\label{sec:data.carbon_corr}
		
		The observed carbon abundances would be affected by CN cycle in the lower layer of atmosphere during red giant branch \citep[e.g.][]{aoki_carbonenhanced_2007}.
		Consequently, $\mathrm{ [(C+N)/Fe]}$ is used to counteract the evolutionary effect \citep[e.g.][]{ishigaki_initial_2018}.
		However, the nitrogen abundance is still difficult to be measured with high precision.
		In this work, an alternative solution is used to estimate surface carbon depletion by applying a carbon correction according to the observed surface gravity $\log g$, metallicity $\mathrm{[Fe/H]}$ and $\mathrm{[C/Fe]}$ \citep{placco_carbon-enhanced_2014}.
		
		In summary, the analysis of the remainder of the paper is based on the abundances of C, Na, Mg, Al, Si, Ca, Ti, Mn, Fe, Co, and Ni. 
		The majority of the abundances, except for C, Na, Mg, and Al, are adopted from the literature which are based on the \ac{LTE} assumption.
		The C abundances used throughout the rest of the work refer to the corrected values according to their stellar evolutionary phase based on \citet{placco_carbon-enhanced_2014} with the assumption of $\mathrm{[N/Fe]}=0$.
		In addition for Na, Mg and Al, the NLTE corrections reported by \citet{lind_non-lte_2022} are adopted and incorporated into the abundance set.

		\section{Method}
		\label{sec:method}
		
		\subsection{Nucleosynthesis Yields of Massive Stars}
		
		Recent decades have witnessed extensive studies concerning zero-metallicity \ac{CCSN}e.
		Various mechanisms have been proposed, including hydrostatic burning and the correction of explosive nucleosynthesis \citep{heger_nucleosynthesis_2010}, mixing-fallback during explosion \citep{umeda_nucleosynthesis_2002}, and the effect of rotation \citep{limongi_presupernova_2012, roberti_zero_2024}. 
		The development of different stellar evolutionary codes, such as KEPLER \citep[e.g.][]{woosley_evolution_2002} and FRANEC \citep[e.g.][]{limongi_presupernova_2012}, has been undertaken.
		For the majority of elements, the discrepancies in nucleosynthesis yields between KEPLER and FRANEC are found to be relatively consistent.
		However, for certain elements such as Sc and Cu, the discrepancies could extend to approximately an order of magnitude \citep{limongi_presupernova_2012}.
		
		In order to ensure the consistency in stellar evolution and element production networks, we only used the nucleosynthesis yields calculated from the same stellar model in \citet{heger_nucleosynthesis_2010}.
		It is because their models are still the most advanced grids to date for zero-metallicity stars, with a minimum mass interval of $0.1\,M_\odot$.
		Three independent initial stellar properties are utilized as input parameters, including \ac{ZAMS} mass $M$, explosion energy $E$ and mixing $\log f_\mathrm{mix}$.
		It is noteworthy that $E$ and $\log f_\mathrm{mix}$ are employed to parameterize the supernova explosion. This parameterized prescription for artificial explosion is necessary due to the current knowledge regarding the explosion mechanism.
		It is inevitable that this will result in the exploration of some unrealistic stellar property spaces and lead to unreliable supernova nucleosynthesis yields.
		The constraints on explosion property spaces are now based on both explodability criterion \citep{zhang_fallback_2008} and abundance observations \citep{jiang_modified_2024}.
		As will be discussed in Section~\ref{sec:discuss.constraint_snmodel}, these two independent methods demonstrate a high degree of consistency in constraining the choice of explosion energy. 
		
		\subsection{Abundance Fitting} 
		\label{sec:method.abufit}
		
		A comparison of the abundance patterns of EMP stars with theoretical nucleosynthesis yields enables one to constrain the properties of the first stars, their supernova explosions, and the subsequent mixing processes \citep{jiang_modified_2024}.
		The discrepancy between observations and theoretical predictions can be quantified using the reduced chi-square statistic, $\chi^2_\nu = \chi^2 / \nu$, where $\nu$ represents the degrees of freedom.
		Automated tools, e.g. STARFIT, perform rapid estimations by selecting the most probable progenitor that minimizes $\chi^2_\nu$ as the solution. 
		
		However, this most probable result derived from minimum $\chi^2$ estimation is challenged by {robustness issues}.
		As demonstrated in a number of studies \citep[e.g.][]{frebel_sd_2015, placco_metal-poor_2015, placco_g64-12_2016, fraser_mass_2017, ishigaki_initial_2018, jiang_modified_2024}, it is importance to take into account the associated uncertainties in progenitor properties.
		Even small fluctuations in observed abundances, within the confines of observational uncertainties, have the potential to result in a significant variation in progenitor mass estimates.
		Given the sensitivity of stellar properties to observed abundances, it is recommended that \ac{probability distributions} should be favored over point estimation in the derivation of progenitors.
		The Monte Carlo method is utilized in abundance fitting to account for the impact of uncertainties from both observations and theoretical models.
		This method involves the simulation of observational effects by independent sampling from normal distributions, with the standard deviations representing observational uncertainties.
		Nevertheless, the inherent uncertainties in supernova yield predictions are often disregarded due to the complexity in quantifying the systematic uncertainties in chemical yields in stellar evolution and explosive nucleosynthesis.
		It is important to acknowledge that the classical Monte Carlo method merely samples \ac{from} independent abundance distributions, which introduces nonphysical biases when interpreting progenitor properties.
		In the context of generating random samples in high-dimensional elemental abundance spaces, it is often the case that a substantial number of sampling iterations are required to ensure the reliability of the results.
		In such circumstances, the Markov chain is typically exploited to facilitate a more efficient exploration of the property space. However, stellar structures and nucleosynthesis yields vary significantly with initial stellar masses, especially at the lower mass end \citep[e.g.][]{heger_nucleosynthesis_2010}. This leads to a high sensitivity of stellar properties to observed abundances. Distributions of progenitor properties with multiple modes due to high sensitivity can cause Markov chain walkers confined to certain regions. In such circumstances, achieving convergence requires an unacceptable number of iteration steps and becomes extremely time-consuming. 
		
		We otherwise recommend exploring the whole property spaces with their likelihood 
		\begin{equation} 
			\label{eqn:likelihood}
			\pi(\mathbf{y}|\bm{\theta}) = \exp(-\chi_\nu^2/2),    
		\end{equation}
		where $\bm{\theta}$ and $\mathbf{y}$ signify the stellar property and observation vectors respectively.
		This shares a similar mathematical essence as the p-value weighted distribution proposed by \citet{ishigaki_initial_2018}. According to Bayesian inference, the resulting \ac{posterior} distribution $p(\mathbf{y}|\bm{\theta})$ could be expressed as:
		\begin{equation} 
			\label{eqn:bayesian}
			p(\bm{\theta}|\mathbf{y}) \propto \pi(\mathbf{y}|\bm{\theta}) p(\bm{\theta})
		\end{equation}
		with a prior $p(\bm{\theta})$ of stellar properties, which {is determined by} the resolution of model grids.
		However, the resulting mass distribution would be biased by the arbitrary choice of distinct property grids adopted in different supernova models.
		This bias will be detailedly discussed in Section~\ref{sec:result.imf_bias}.
		In brief, we recommend to utilize in this work a uniform prior distribution of
		\begin{equation}
			p(\bm{\theta}) \propto \Delta\bm{\theta} = \Delta \log M \Delta \log E \Delta \log f_\mathrm{mix},
			\label{eqn:uniform_prior}
		\end{equation}
		where $\Delta\bm{\theta}$ denotes the resolution of properties. Additionally, this Bayesian inference also allows for incorporation of explodability constraint on stellar properties. 
		
		\section{Result} 
		\label{sec:result}
		
		\subsection{Which Element is More Important?}
		\label{sec:result.element}
		
		It is extensively studied that the derived properties of the progenitors of \ac{EMP} stars would be substantially influenced by the measurement of elemental abundances as mentioned in Section~\ref{sec:method.abufit}. However, the influence of individual element on progenitor derivation is not well understood. \citet{hartwig_machine_2023} proposed that [C/Mg] and [Ca/Fe] are two important abundance ratios in determining the multiplicity of chemical enrichment of EMP stars. In this section, we introduce an innovative method to define the response of progenitor derivation to each element.
		
		\subsubsection{Sensitivity}
		\label{sec:result.element.sensitivity}
		
		\begin{figure*}
			\centering
			\includegraphics[width=.95\linewidth]{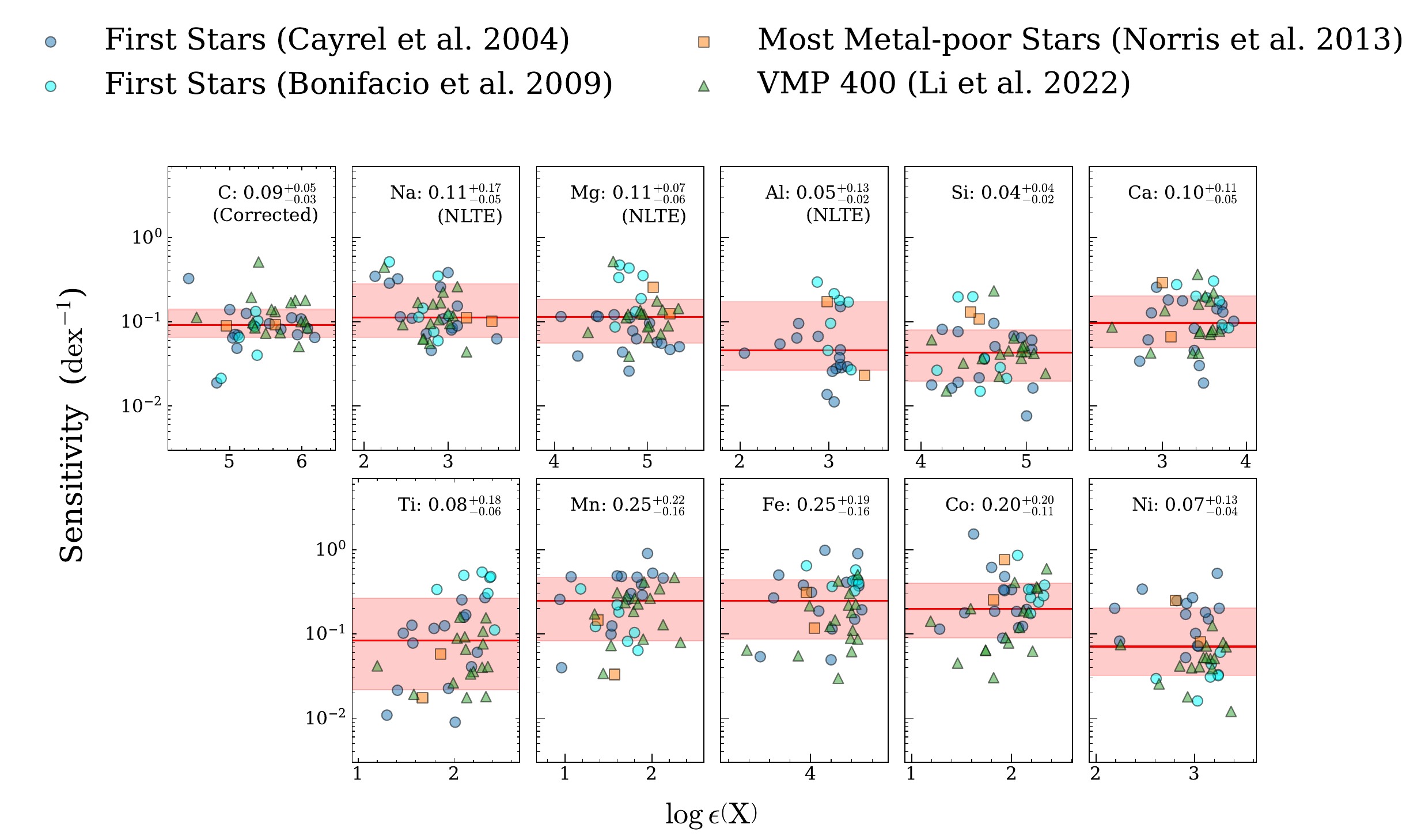}
			\caption{
				The sensitivity of progenitor mass derivation for different elements. 
				In each panel, the scattered points correspond to individual EMP stars from different samples, while the horizontal solid line and shaded regions indicate the 50th, 16th and 84th percentiles of the sensitivities.
				The sensitivity is expressed in the unit of $\mathrm{dex}^{-1}$. 
			}
			\label{fig:sensitivity}
		\end{figure*}
		
		As for the derivation of progenitors, we perturb the abundance of a certain element $X$ and measure the corresponding variation in the properties of the derived progenitors with respect to the elemental abundance in Wasserstein metric, 
		\begin{equation}\label{eqn:derivative}
			\frac{1}{\delta{\mathbf{y}_X}}
			\int \left| P(\bm{\theta}|\mathbf{y}) - P(\bm{\theta}|\mathbf{y}+\delta{\mathbf{y}_X}) \right| \mathrm{d}\bm{\theta}, 
		\end{equation}
		where $P(\bm{\theta}|\mathbf{y})$ signifies the \ac{CDF} of the posterior $p(\bm{\theta}|\mathbf{y})$, i.e.
		\begin{equation}\label{eqn:cumulative_density}
			P(\bm{\theta}|\mathbf{y}) = \int p(\bm{\theta}|\mathbf{y})\mathrm{d}\bm{\theta}.
		\end{equation}
		To focus on the progenitor mass $M/M_\odot$, other properties, including energy $E$ and mixing parameter $\log f_\mathrm{mix}$ are marginalized.
		For the integration, we use $\log M/M_\odot$ rather than $M/M_\odot$ itself, so that relative uncertainties across different mass ranges are taken into account properly.
		Consequently, the sensitivity is expressed in the unit of $\mathrm{dex}^{-1}$ to denote the change in the logarithm scale of progenitor mass per unit of observational uncertainty.
		The value of Equation~\ref{eqn:derivative} is hereafter referred to as the sensitivity of element $i$, and serves as a measure of the reliability of the progenitor derivation.
		It should be noted that the sensitivity is defined over the entire grid, reflecting the fact that the inferred progenitor mass distribution of each star may shift due to the uncertainties or systematics in observational abundances.
		This parameter provides a comprehensive assessment of how the observational abundance of each element affect the progenitor‐mass distribution: 
		it not only describes the change in the most probable progenitor mass caused by observational uncertainties, 
		but also captures the corresponding difference in broadening (or dispersion) of the distribution.
		
		In cases where no observational uncertainty is available, we adopt a representative value of $0.1\,\mathrm{dex}$. 
		Since the quantity of sensitivity in Equation~\ref{eqn:derivative} is defined as the derivative of the posterior with respect to the observed abundances, 
		it is not expected to exhibit a strong dependence on the adopted value assumed for the observational error. 
		Specifically, we also test with a larger uncertainty. 
		It only produces a negligible change in the sensitivity parameter, further confirming the robustness of its definition.
		Accordingly, the y-axis values in Figure~\ref{fig:sensitivity} can be viewed as the uncertainty in the derived mass inevitably caused by observations. 
		In addition to the potential influence of the adoption of different observational errors, it is also relevant to consider whether the sensitivities are affected by the underlying abundance distributions of the elements.
		For instance, as illustrated in Figure~\ref{fig:abudist}, carbon shows substantially larger scattering than most other elements and its distribution differs across samples.
		Stars in the FS sample generally display higher carbon abundances than those in the VMP 400 sample.
		However, as demonstrated in Figure~\ref{fig:sensitivity}, there appears to be no clear correlation between the sensitivities and the abundance for each element, indicating that the sensitivity results are not primarily driven by sample-dependent abundance distributions.
		
		In this work, the sensitivity is defined as the effect per unit of abundance uncertainty, which reflects an intrinsic property of each element in progenitor derivation
		Therefore, the effective influence of a specific element should be estimated by multiplying its intrinsic sensitivity by the corresponding observational uncertainties.
		Given that the typical observational precision for elemental abundances in metal-poor stars is around $0.1\,\mathrm{dex}$, the sensitivity in our Bayesian framework remains at the level of $\sim0.01$. 
		This indicates that the inferred progenitor properties are relatively stable, corresponding to only about $2\%$ uncertainty in stellar mass.
		Nevertheless, the sensitivities still vary across different elements.
		In particular, Mn, Fe, and Co exhibit stronger influences, suggesting that iron-peak elements play a more critical role in constraining the progenitors of EMP stars.
		Furthermore, the observational uncertainties in different elements vary due to current measurement limitation. 
		Specifically, typical uncertainties for most elements are approximately $0.1\,\mathrm{dex}$, whereas those for C, Mg and Al are comparatively larger at around $0.2\,\mathrm{dex}$. 
		Consequently, while Mn, Fe and Co exert a pronounced influence on progenitor derivation primarily due to their inherent sensitivities, C and Mg also contribute substantially, largely attributable to their augmented observational uncertainties.
		For these elements, the resulting effect corresponding to about a $4\%$ variation in the inferred progenitor mass.
		
		\subsubsection{Residuals of Abundances between Theoretical Predictions \& Observations}
		\label{sec:result.element.residual}
		
		\begin{figure}
			\centering
			\includegraphics[width=.8\linewidth]{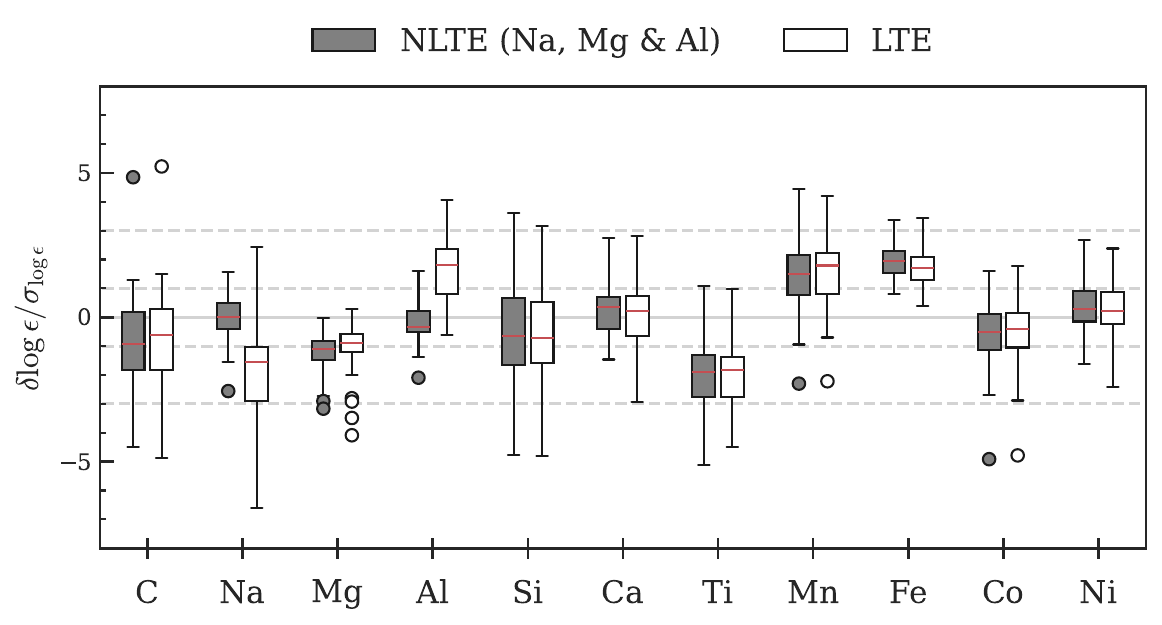}
			\caption{
				Distribution of the abundance residuals between predictions and observations over their corresponding observational uncertainties. 
				The left (gray) and right (white) boxes represent the NLTE-corrected and LTE abundances from literature.
				Each box extends from the first to the third quartile of the corresponding abundance residuals, with an red solid line indicating the median value. 
				Whiskers extend to the furthest abundance residuals lying within $1.5\times$ the \ac{IQR}s from the boxes. 
				The dashed horizontal lines signify the deviations of $1 \sigma$ and $3 \sigma$ of observational uncertainty.
			}
			\label{fig:discpreancy_theory_observ}
		\end{figure}
		
		In addition to assessing the sensitivity analysis, we also need to understand the deviation between the observed abundances and theoretical prediction. 
		Such deviation may result from limitations in the nucleosynthesis theories, observational process, systematic effects or a combination of these factors. 
		To investigate this, we treat the element of interest as non-detection and predict its abundance based on the remaining observed elements. 
		The predicted abundance is inferred from the aforementioned Bayesian framework, where the estimated abundance is obtained by integrating the posterior probability, as defined in Equation~\ref{eqn:derivative}. 
		To further explore the impact of different assumptions, both LTE and NLTE abundances of Na, Mg and Al are examined, and the resulting residual distributions of different elements are summarized in Figure~\ref{fig:discpreancy_theory_observ} in the form of boxplots, where each residual is normalized by its corresponding observational uncertainty.
		
		As shown in Figure~\ref{fig:discpreancy_theory_observ}, our predicted elemental abundances generally fall within $3\sigma$ of the observational uncertainty, although different elements exhibit distinct behaviors. 
		This section will commence with a comprehensive examination of the impact of \ac{NLTE} effects on residuals.
		As the L22 \ac{NLTE} corrections are available for Na, Mg and Al, the present discussion is limited to these elements.
		Our analysis in Section~\ref{sec:data.nlte.namgal} indicates that odd-Z elements, i.e. Na and Al, are both strongly affected by \ac{NLTE} effects.
		The application of \ac{NLTE} corrections leads to a substantial reduction in the residuals of Na and Al, along with a notable decrease in their scattering, as illustrated in Figure~\ref{fig:discpreancy_theory_observ}.
		This finding suggests that the residuals of Na and Al can be largely attributable to the \ac{NLTE} effect.
		This highlights the importance of accounting for \ac{NLTE} effects when interpreting the abundances of odd-Z elements. 
		
		Nevertheless, even after applying \ac{NLTE} corrections, a systematic underestimation of Mg by approximately $1\sigma$ $(-0.2\,\mathrm{dex})$ persists even after \ac{NLTE} corrections. 
		This indicates that the actual observed Mg abundances are slightly higher than those predicted by theoretical yields. 
		Given that massive stars are considered the primary source of Mg in \ac{EMP} stars in the early Universe, this residual may suggest an underestimation in Mg production in current nucleosynthesis yield models. 
		However, Mg enhancements could also arise from complex enrichment processes, such as inhomogeneous mixing in multiple \ac{CCSN}e events. 
		There is also a possibility of rapid chemical enrichment from second-generation massive stars in the early Universe, as suggested by recent zoom-in cosmological simulations \citep{mead_aeos_2025}. 
		
		Significant residuals are also observed in Ti, Mn and Fe, typically exceeding one observational uncertainties.
		The \ac{NLTE} corrections of Na, Mg, and Al have a negligible influence on the predicted abundances of other elements.
		As shown in Figure~\ref{fig:discpreancy_theory_observ}, elements without \ac{NLTE} corrections exhibit no significant change in their residuals.
		This minimal influence also verifies the robustness and reliability of the analysis for other elements. 
		In the absence of corrections for \ac{NLTE} effects for these elements in the present analysis, it still remains inconclusive whether their residuals arise from observational effects or limitations in theoretical yields.

		\subsubsection{The Influence of Non-detection}
		\label{sec:result.element.nondetection}
				
		\begin{figure*}
			\centering
			\includegraphics[width=.95\linewidth]{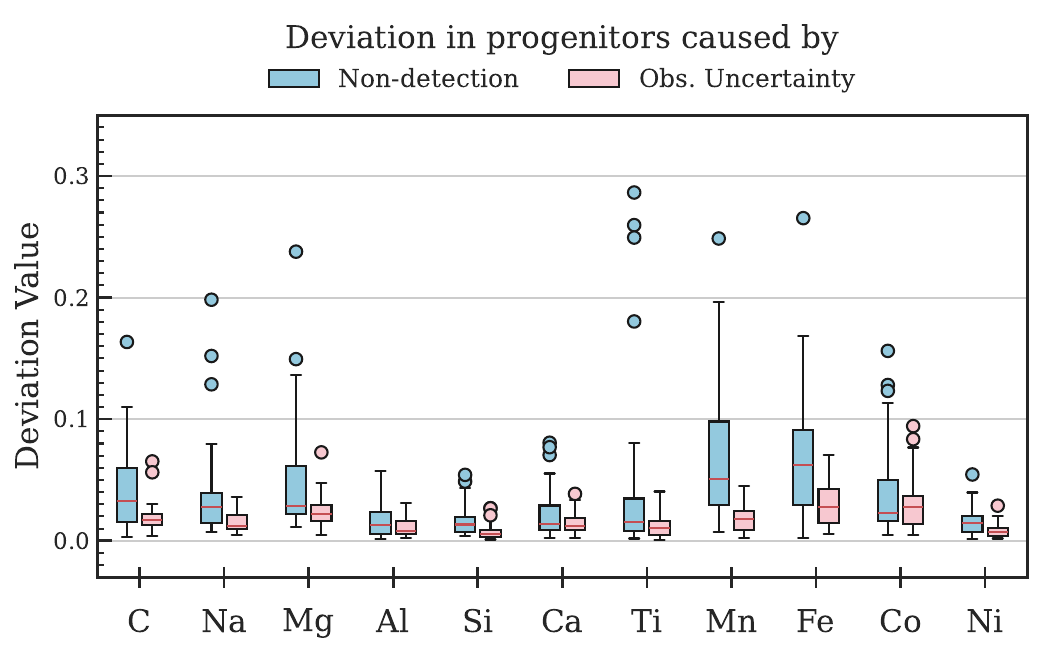}
			\caption{
				Deviation in progenitor mass caused by the non-detection (blue, left) and observational uncertainty (red, right) of each element, summarized in boxplots. The definition of boxes and whiskers follows that in Figure~\ref{fig:discpreancy_theory_observ}.
				NLTE abundances are adopted for Na, Mg, and Al in this analysis.
			}
			\label{fig:nondetection}
		\end{figure*}
		
		Due to observational limitations such as spectral coverage or signal-to-noise ratio, abundance measurements of certain elements in \ac{EMP} stars are occasionally unavailable. 
		To account for this, further evaluation is necessary to ascertain the impact of such non-detection on progenitor derivation.
		In such cases, the absent element contributes no information to the inference. 
		In the abundance fitting procedure, this is implemented by setting its $\chi^2$ contribution to zero. 
		Alternatively, this is also analogous to adopting the predicted abundance predicted from the remaining measured elements.
		As outlined in the preceding section, the majority of residuals between predictions and observations fall within $3\sigma$ of the observational uncertainty.
		However, the absence of individual elements may still introduce a non-negligible effect. 
		To quantify this, for an element $X$ with no available abundance measurement, a 1D Wasserstein metric analogous to Equation~\ref{eqn:derivative} is adopted:
		\begin{equation}\label{eqn:nondetect}
			\int \left| P(\log M/M_\odot|\mathbf{y}) - P_X(\log M/M_\odot|\mathbf{y}) \right| \mathrm{d}\log M/M_\odot, 
		\end{equation}
		where $P$ and $P_X$ represent the \ac{CDF} of posterior with and without the abundance of element $X$, respectively. 
		For comparison, the deviations arising from non-detection and those induced by observational uncertainties, are summarized in Figure~\ref{fig:nondetection}. 
		The deviations due to observational uncertainties are calculated by the integral term in Equation~\ref{eqn:derivative}, i.e. the product of the sensitivity and the corresponding $\delta{\mathbf{y}_X}$.

		A comparison between the effects of non-detection and observational uncertainty reveals that the non-detection generally produces a significantly larger deviation in inferred progenitor mass, with an average deviation of approximately $0.02$, corresponding to about $4\%$ in stellar mass.
		As illustrated by the $1\sigma$ reference line in Figure~\ref{fig:discpreancy_theory_observ}, the residuals between the predicted and observed abundances commonly exceed their associated observational uncertainties.  
		Such significant residuals would amplify the impact of non-detection, thus rendering the measurement of these elements particularly crucial.
		For instance, Fe and Mn (with mean deviations of $0.06$ and $0.05$, respectively), followed by C and Mg (both with mean deviations of $0.03$), exhibit the most significant influence on progenitor mass inference. 
		This underscores the significance of precise detection of critical elements. It should be noted, however, that the consequences of non-detection are not entirely determined by residual.
		For elements such as Na, Al and Ni, whose residuals are typically below $1\sigma$, non-detection also exerts a more significant influence than uncertainty.
		
		\subsection{IMF \& Resolution Bias} 
		\label{sec:result.imf_bias}
		
		The \ac{IMF} derived from the abundance patterns of \ac{EMP} stars has been investigated in multiple studies.
		In this context, we present in Table~\ref{tab:imf_indivi} the resulting progenitor mass distribution for individual \ac{EMP} stars based on our Bayesian analysis.
		
		\startlongtable
		\begin{deluxetable*}{lcccccccccc}
			\tablecaption{Progenitor Mass Distribution from Bayesian Abundance Fitting}
			\tablehead{
				Object name  & $P\left(M=  9.6\,M_\odot\right)$ & $P\left(M=  9.7\,M_\odot\right)$ & $P\left(M=  9.8\,M_\odot\right)$ & $P\left(M=  9.9\,M_\odot\right)$ & $P\left(M= 10.0\,M_\odot\right)$ & ...
			}
			\startdata
			BD-18 5550   & 0.00E+00 & 0.00E+00 & 4.70E$-$07 & 4.20E$-$05 & 6.96E$-$10 & ... \\
			BS 16477-003 & 0.00E+00 & 0.00E+00 & 3.44E$-$04 & 1.46E$-$03 & 1.48E$-$06 & ... \\
			CS 22172-002 & 0.00E+00 & 0.00E+00 & 4.82E$-$14 & 2.61E$-$06 & 4.33E$-$19 & ... \\
			CS 22186-025 & 0.00E+00 & 0.00E+00 & 2.09E$-$13 & 5.48E$-$05 & 6.38E$-$15 & ... 
			\enddata
			\label{tab:imf_indivi}
			\tablecomments{Table~\ref{tab:imf_indivi} is published in its entirety in the machine-readable format. A portion is shown here for guidance regarding its form and content.}
		\end{deluxetable*}
		
\begin{figure}
			\centering
			\includegraphics[width=.75\linewidth]{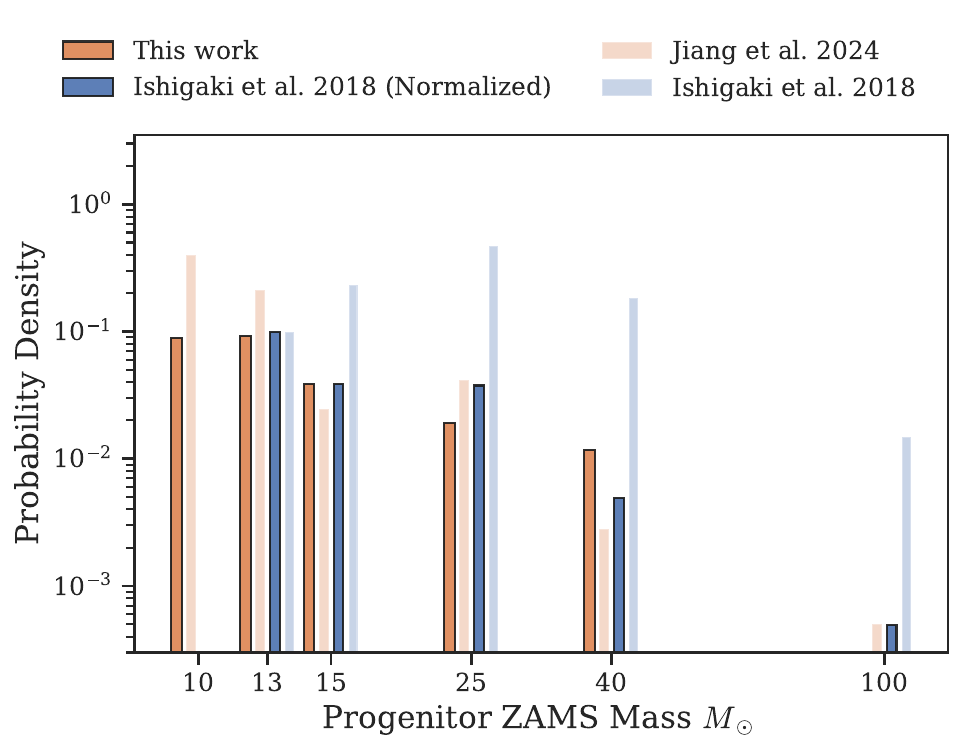}
			\caption{
				Comparison between the resolution-normalized results. The first (orange), second (translucent orange), third (blue) and right (translucent blue) bars respectively represent the resulting \ac{IMF}s of this work and \citet{jiang_modified_2024}, the normalized and raw results of \citet{ishigaki_initial_2018}. 
			}
			\label{fig:reappearance}
		\end{figure}

		However, systematic biases still exist between these studies due to different supernova models and their adopted mass grids.
		This bias leads to different functional forms of \ac{IMF}, including a power-law distribution \citep{fraser_mass_2017, jiang_modified_2024} or log-normal distribution \citep{ishigaki_initial_2018}. 
		We adopt the uniform prior defined in Equation~\ref{eqn:uniform_prior} to normalize results derived under different resolutions.
		The normalized outcomes are then compared with the p-value weighted result from \citet{ishigaki_initial_2018} and the mono-enriched result from \citet{jiang_modified_2024}. 
		To ensure consistency in the parameter ranges of \ac{ZAMS} mass $M$, explosion energy $E$ and mixing $\log f_\mathrm{mix}$ across different supernova models, we apply a simplified filter to the \citet{heger_nucleosynthesis_2010} models based on the parameter space used by \citet{ishigaki_initial_2018}.
		In particular, we exclude low-mass hypernovae, i.e. models with $M<20\,M_\odot$ with $E\geq5\,B\left(10^{51}\,\mathrm{erg}\right)$.
		Moreover, since \citet{ishigaki_initial_2018} does not include massive stars with $M<13\,M_\odot$, all derived \ac{IMF}s are normalized such that the total probability for $M>13\,M_\odot$ satisfies $p\left(M>13\,M_\odot\right)=1$, to facilitate visual comparability in Figure~\ref{fig:reappearance}.
		As shown in Figure~\ref{fig:reappearance}, the discrepancies arising from varying grid resolutions are significantly reduced. 
		Moreover, all normalized results exhibit a well-defined power-law trend and an overdensity in the $13-15\,M_\odot$ mass bin.
		In addition, a secondary overdensity is also present at $30-40\,M_\odot$, as shown in Figure~\ref{fig:fit_imf_edf}.
		
		\section{Discussion} 
		\label{sec:discuss}
		
		\subsection{Observational Constraints on Supernova Models}
		\label{sec:discuss.constraint_snmodel}
		
		The chemical signatures of EMP stars provide reliable alternatives to constraining the nature of Pop III stars and their supernova explosions at the easiest access.
		In this section, we present a detailed analysis of the two-dimensional distribution of \ac{ZAMS} mass and explosion energy, which is calculated with the same exclusion criterion adopted in Section~\ref{sec:result.imf_bias}, namely the removal of models with $M<20\,M_\odot$ and $E \geq 5 \times 10^{51}\,\mathrm{erg}$. 
		As illustrated in Figure~\ref{fig:2d_progen_dist}, this distribution enables quantitative constraints on the \ac{IMF} of the first stars and their supernova mechanisms.
		
		\subsubsection{Explodability}
		\label{sec:discuss.constraint_snmodel.explodability}
		
		According to \citet{zhang_fallback_2008}, massive stars successfully explode only the mass of their remnants falls beneath the limiting baryonic mass $M_\mathrm{bary}$, which is given by
		\begin{equation}
			M_\mathrm{bary} = M_\mathrm{grav}\left[1-7.38\times10^{-2}\left(M_\mathrm{grav}/M_\odot\right)\right]^{-1}, 
		\end{equation}
		where $M_\mathrm{grav}$ denotes the maximum gravitational mass of neutron stars. 
		Recent multi-messenger observation of neutron star merger \citep{margalit_constraining_2017} suggests a maximum gravitational mass of $M_\mathrm{grav} = 2.17,M_\odot$ at the $90\%$ confidence, which corresponds to a limiting baryonic mass of $2.58\,M_\odot$. 
		Based on this observational limit, we update the explodability constraints from \citet{zhang_fallback_2008}, as shown by the hatched region in Figure~\ref{fig:2d_progen_dist}. 
		
		Observational constraints play a crucial role in refining supernova models. In \citet{jiang_modified_2024}, the derived distribution of progenitor properties incorporates supernova explodability, under the assumption that the metal enrichment of \ac{EMP} stars are attributed to the explosion of first \ac{CCSN}e. An alternative approach to constraining realistic supernova models involves minimizing reduced chi-squared $\chi^2_\mathrm{red}$ of each model relative to the observed abundance patterns of \ac{EMP} stars \citep{hartwig_machine_2023}. In our Bayesian inference framework, this $\chi^2_\mathrm{red}$ minimum criterion is inherently incorporated and reflected as the extremely low posterior probabilities in the resulting progenitor distribution, which are coded with black in Figure~\ref{fig:2d_progen_dist}. 
		All models with posterior probabilities below $p(\bm{\theta}|\mathbf{y})<10^{-3}$  represented in black.
		
		Notably, the theoretically predicted non-exploding models coincide with those exhibiting extremely low posterior probabilities, specifically those satisfying $p(\bm{\theta}|\mathbf{y}) < \max p(\bm{\theta}|\mathbf{y})/100$. Therefore the theoretical constraint on the combination of mass and energy is consistent with the observational constraint. 
		
		\subsubsection{Mass--energy Relation}
		\label{sec:discuss.constraint_snmodel.mer}
		
\begin{figure*}
			\centering
			\includegraphics[width=1.\linewidth]{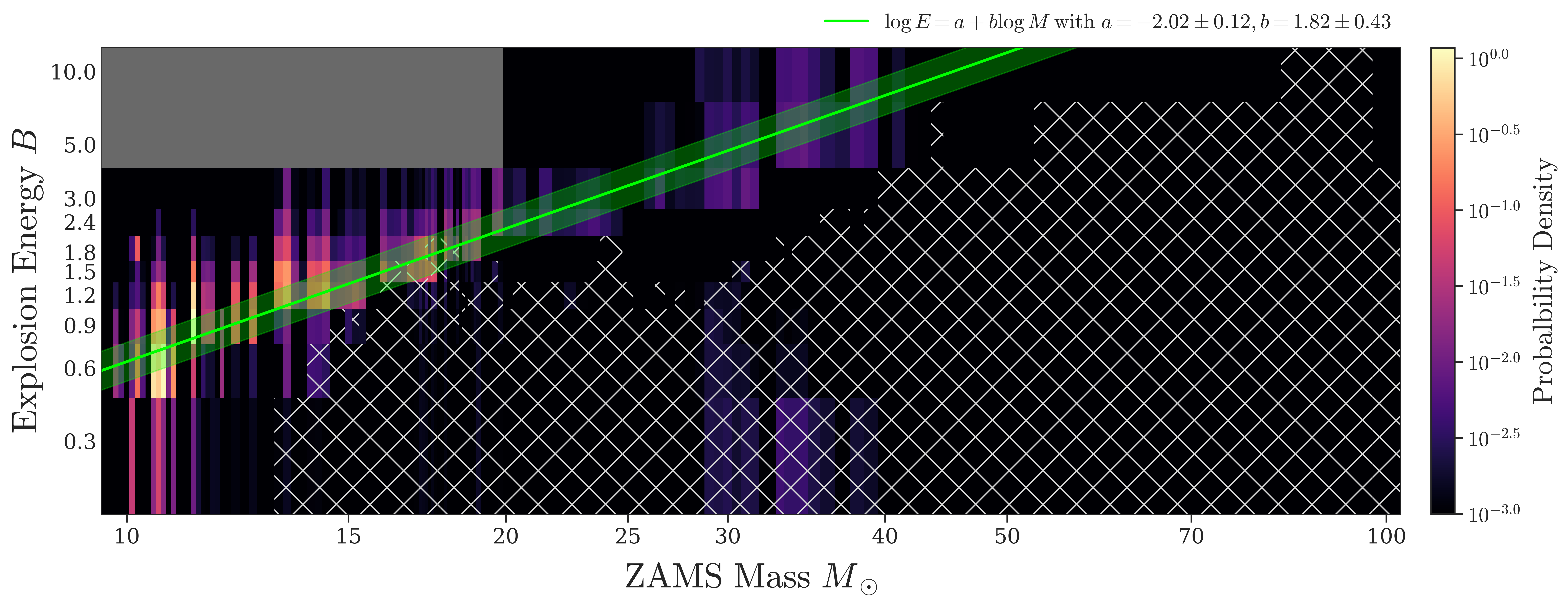}
			\caption{
				Progenitor distribution derived from abundances of \ac{EMP} stars and explodability constraints inferred from neutron star mass maximum. The probability densities of progenitor models are color-coded following the colorbar at right panel, with models having posterior probabilities below $10^{-3}$ shown in black. Non-exploding models are indicated by region hatched with cross lines, and pre-excluded models are shaded in gray.
			}
			\label{fig:2d_progen_dist}
		\end{figure*}

		An evident linear correlation between mass and energy can be detected in Figure~\ref{fig:2d_progen_dist}, with both the x- and y-axes plotted on logarithmic scales.
		This trend supports a relationship between progenitor mass and explosion energy: $\log E = a + b \log M$.  
		To quantitatively evaluate this relation, we perform a linear regression in $\log$-–$\log$ space using the code \texttt{LtsFit} \citep{cappellari_atlas3d_2013} by sampling progenitors following the resulting two-dimensional distribution. 
		The uncertainties in $\log E$ and $\log M$ are set to $0.1\,\mathrm{dex}$ and $0.5\,\mathrm{dex}$ respectively, which are roughly consistent with the kernel widths derived in Section~\ref{sec:discuss.imf_edf_fit}.
		The regression yields best-fit parameters $a = -2.02\pm0.12$ and $b = 1.82\pm0.43$. The resulting relation and its associated $\pm1\sigma$ confidence region are shown in Figure~\ref{fig:2d_progen_dist} as a solid green line and a surrounding transparent green band, respectively.
		The Spearman and Pearson correlation coefficients are $0.7$ and $0.8$ respectively, both with $p$-value much less than $0.01$.
		This indicates with high statistical significance that the first supernovae roughly follow a mass--energy relation of
		\begin{equation}
			E \propto M^{2}.
			\label{eqn:mass_energy_relation}
		\end{equation}
		
		This finding is fully consistent with the relation between kinetic energy and initial mass of $E \propto M^2$ proposed in Figure~4 of \citet{poznanski_emerging_2013}. It also aligns well with a recent theoretical result from \citet{fang__inferring_2023}. They report a relation of $E \propto M_\mathrm{CO}^{1.3}$ and $M_\mathrm{CO} \propto M^{1.5}$, implying $E \propto M^{1.95}$—very close to our result. Notably, since these studies are primarily based on solar-metallicity progenitors, the agreement suggests that the mass--energy relation holds for both zero and solar metallicity environment. Whether this relation is universal remains an open question requiring further investigation at intermediate metallicities.
		These results are limited by the relatively small size of the homogeneous \ac{EMP} star samples.
		Continued observational efforts to expand the sample will be crucial to further test and validate this trend.
		A statistically larger sample will allow for more robust constraints on the mass-energy relation of the theoretical supernova models.
		
		\subsection{Modified IMF \& EDF with Explodability}
		\label{sec:discuss.imf_edf_fit}
		
		For massive stars, it is commonly hypothesized that both the \ac{ZAMS} mass and explosion energy follow power-law distributions.
		The \ac{IMF} and \ac{EDF} could be written by denoting their respective exponents as $\alpha_m$ and $\alpha_e$:
		\begin{eqnarray}
			p(M) &\propto& M^{-\alpha_m},\\
			p(E) &\propto& E^{-\alpha_e}.
			\label{eqn:imf&edf}
		\end{eqnarray}
		In previous studies \citep[e.g.][]{koutsouridou_energy_2023, jiang_modified_2024}, the mass and energy are often assumed to be independent to widen the supernova model parameter space.
		However, this can inevitably introduce unreliable nucleosynthesis yields in both progenitor derivation and chemical evolution modeling \citep{hartwig_machine_2023}.
		Based on the inferred mass--energy relation of Equation~\ref{eqn:mass_energy_relation},  the independence assumption no longer holds.
		Applying the transformation method of probability, the exponents are related by:
		\begin{equation}
			2\alpha_e=\alpha_m+1.
			\label{eqn:exponent_relation}
		\end{equation}
		
		However, since the metal abundances observed in \ac{EMP} stars trace only the chemical yields of successfully exploding massive stars, the derived mass distribution must be corrected for explodability.
		Following Equation~19 of \citet{jiang_modified_2024} and breaking the independence assumption, the modified \ac{IMF} becomes
		\begin{equation}
			p(M) \propto \zeta(M)M^{-\alpha_m},
			\label{eqn:modified_imf}
		\end{equation}
		where $\zeta(M)$ is set to $1$ if the massive star explodes successfully and $0$ otherwise.
		The \ac{EDF} is modified similarly:
		\begin{equation}
			p(E) \propto \zeta(M(E))E^{-\alpha_e}.
			\label{eqn:modified_edf}
		\end{equation}
		Although significant effort has been devoted to determining the explosion mechanism of \ac{CCSN}e, the precise explodability boundary across the full mass spectrum remains poorly constrained.
		For instance, when compactness is used as a proxy for explodability \citep{oconnor_black_2011}, non-negligible variations in predictions are found across different stellar evolution models \citep[e.g.][]{chieffi_presupernova_2020}. 
		Nonetheless, a non-monotonic dependence of explodability with the initial mass is consistently reported.
		While the use of compactness as a predictor of explodability has been debated \citep{burrows_core-collapse_2021} and alternative indicators have been also proposed \citep[e.g.][]{ertl_two-parameter_2016, boccioli_explosion_2023}, a non-monotonic dependence of explodability on initial mass has been repeatedly reported. 
		
\begin{figure}
			\includegraphics[width=1.\linewidth]{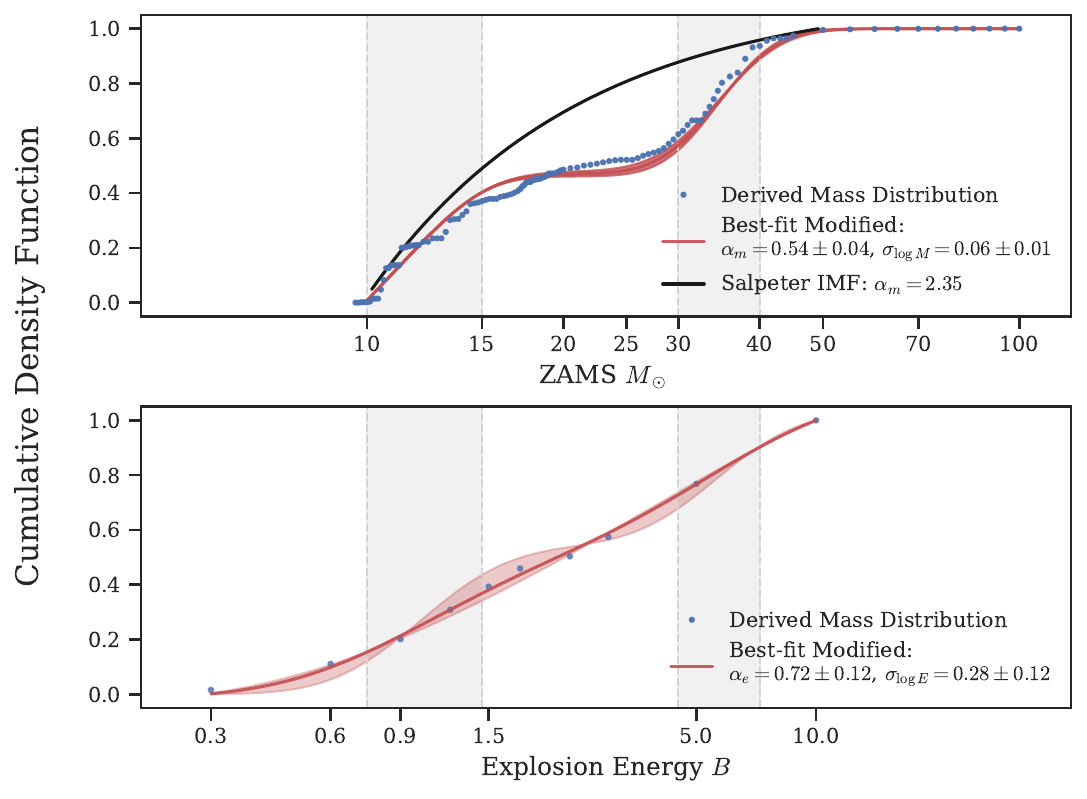}
			\caption{
				\ac{IMF} (upper panel) and \ac{EDF} (bottom panel) of the first stars in \ac{CDF}. The blue dots represent the derived \ac{CDF} of mass and energy distribution from the \ac{EMP} star sample, while the red lines with shaded regions standards for their corresponding best-fit distribution as defined in Equation~\ref{eqn:broaden_modified_imf_edf}. In each panel, the shaded gray zones are the assumed successfully explosive models from concentrated intervals in the derived mass distribution. The Salpeter \ac{IMF} \citep{salpeter_luminosity_1955} is coded with black for comparison.
			}
			\label{fig:fit_imf_edf}
		\end{figure}

		Consequently, fitting the derived \ac{IMF} becomes challenging due to this degeneracy between the \ac{IMF} exponent and the explodability. 
		We can only infer possible locations of explodability transitions from features in the derived progenitor mass distribution.
		As shown in Figure~\ref{fig:fit_imf_edf}, the derived mass distribution is concentrated in two distinct regions: one with $M<15\,M_\odot$, and the other with $30\,M_\odot<M<35\,M_\odot$. 
		Together, these two intervals contain approximately $68\%$ of the total probability. 
		We therefore interpret them as ``islands of explodability" in initial mass \citep{sukhbold_compactness_2014}. 
		This interpretation is supported by numerical simulations: \citet{ebinger_pushing_2020} predict successful explosions of stars with $M<15\,M_\odot$, and their associated explosion properties are consistent with our mass--energy relation. 
		However, the explosion energies for stars above $15\,M_\odot$ in their models are lower than those implied by our relation, making their explodability predictions inapplicable under this scenario.
		Although the detailed pattern of explodability across the mass spectrum is not fully understood, we adopt a simplified assumption: 
		\begin{equation}
			\zeta(M)=\left\{
			\begin{array}{rcl}
				1& &10\,M_\odot<M<15\,M_\odot\ or\ 30\,M_\odot<M<40\,M_\odot,\\
				0& &15\,M_\odot<M<30\,M_\odot\ or\ 40\,M_\odot<M.
			\end{array}
			\right.
			\label{eqn:broaden_modified_imf_edf}
		\end{equation}
		
		Assuming power-law distributions, this study fit the \ac{IMF} and \ac{EDF} in the $\log$ space to obtain the optimal parameters of $\alpha_m$ and $\alpha_e$. 
		The derived distribution, however, is broadened by several sources of uncertainty, including limited sample size and the propagation of observational uncertainties from abundance measurements. 
		The sample size is limited because we utilize only homogeneous \ac{EMP} star samples, which improves the consistency in spectroscopic observations and abundance measurements.
		In addition, the resolution of parameter grids in the adopted supernova models also contributes to this broadening. 
		Specifically, the coarser grids of explosion energy would produce a broader distribution compared to that for \ac{ZAMS} mass. 
		In order to properly address the influence of broadening, we apply a Gaussian kernel convolution to the both modified \ac{IMF} and \ac{EDF} in Equation~\ref{eqn:modified_imf} and~\ref{eqn:modified_edf},
		\begin{eqnarray}
			\tilde{p}(\log M)	&=& p(\log M)*G(\log M; \sigma_{\log M}), \label{eqn:gauss_conv_modimf}\\
			\tilde{p}(\log E)	&=& p(\log E)*G(\log E; \sigma_{\log E}), \label{eqn:gauss_conv_modedf}
		\end{eqnarray}
		where the kernel widths $\sigma_{\log M}$ and $\sigma_{\log E}$ are treated as free-fitted parameters to model the broadening effect. 
		
		To mitigate binning effects, we fit the broadened, explodability-modified \ac{IMF} and \ac{EDF} (in cumulative form) to the distributions derived from the EMP star abundance patterns. 
		The best-fit exponents are $\alpha_m=0.54$ and $\alpha_e=0.72$, which is roughly consistent with the exponent relation given in Equation~\ref{eqn:exponent_relation}.
		These exponents differ from those obtained under the assumption of independence between mass and energy in \citet{jiang_modified_2024}.
		The best-fit kernel widths are $\sigma_{\log M} = 0.06$ and $\sigma_{\log E} = 0.28$, consistent with expectations based on the resolutions of supernova model grids and the propagation of observational uncertainties.
		This further supports the credibility of our Bayesian framework and modified, broadened \ac{IMF} and \ac{EDF}. 
		We note that the inferred \ac{IMF} and \ac{EDF} depend on the assumptions regarding supernova explodability and may vary with different assumptions.
		Subsequent enhancements to the explodability and supernova mechanism will facilitate the refinement of these results.
		
		\section{Conclusion}
		
		This study systematically investigates the explosion mechanisms and progenitor properties of Pop III supernovae by integrating Bayesian inference with the detailed analyses of abundance patterns of extremely metal-poor (EMP) stars.
		We compiled three homogeneous samples of EMP stars from high-resolution spectroscopy and applied the most recent NLTE correction from \citet[][L22]{lind_non-lte_2022} on Na, Mg and Al abundances.
		Our in-depth assessment quantitatively reveals that the discrepancies in abundances could arise from both the atmospheric parameter determination and curve-of-growths implemented in different atmospheric models. 
		
		We also evaluated the influence of individual elements on progenitor derivation. 
		In this study, we propose a novel quantitative metric, "sensitivity", which quantifies how the uncertainty in the abundance of a specific element affects progenitor inference.
		Within the Bayesian framework, the mass shift induced by typical observational uncertainties remains modest.
		The sensitivity metric can be interpreted as indicating that observational uncertainties introduce, on average, an uncertainty of approximately $2\%$ in the inferred progenitor mass.
		Element, C, Mg, Mn, Fe and Co exhibit relatively larger impacts, with mean deviations of around $4\%$.
		For C and Mg, this is primarily driven by their comparatively larger observational uncertainties under current measurement limitations.
		
		Residuals between observed abundances and theoretical yields are computed within the Bayesian framework under both LTE and NLTE assumptions. 
		The robustness of this residual analysis is supported by the negligible differences found for elements without NLTE correction.
		The NLTE effect plays a critical role in interpreting the residuals of odd-Z elements. 
		As shown in Figure~\ref{fig:discpreancy_theory_observ}, NLTE corrections significantly reduce the discrepancies for Na and Al.
		Nonetheless, due to the limited elemental coverage of the L22 grid, which does not include species such as Ti, Mn and Fe, the large residuals observed for these elements remain inconclusive.
		The discrepancy between theoretical predictions and observations for Ti has long remained an unresolved issue in galactic chemical evolution \citep{kobayashi_origin_2020}.
		Future NLTE corrections across a broader range of elements, performed within consistent modeling frameworks, will be essential to distinguish whether the remaining residuals arise from NLTE effects or shortcomings in current nucleosynthesis models. 
		
		In addition, due to observational limitations, including wavelength coverage and signal-to-noise ratio, it might be challenging to obtain complete abundance measurements for all eleven selected elements for some stars. 
		Therefore, an analogous method is employed to analyze the influence of non-detection.
		The findings of the present study reveal that the impact of non-detection on progenitor mass is considerably stronger than that of observational uncertainties, often by a factor of $\sim2$. 
		For certain elements of significance, such as Mn and Fe, the absence of abundance measurements could lead to substantial deviations in progenitor mass, with deviations exceeding $0.05$, corresponding to $>10\%$ in progenitor mass.
		This effect of non-detection is especially pronounced for elements with large residuals in prediction and observation, as these elements contribute significantly to $\chi^2$ statistic in abundance fitting.
		
		We construct the two-dimensional probability distribution of zero-age main sequence (ZAMS) mass and explosion energy of the EMP progenitors, incorporating observational uncertainties in the elemental abundances. 
		Nonphysical models with low ZAMS mass but high explosion energy are excluded a priori. 
		Our Bayesian approach mitigates systematic biases from differences in supernova model grid resolution by adopting a uniform prior on progenitor properties (Section~\ref{sec:result.imf_bias}). 
		Linear regression in $\log$--$\log$ space on the resulting progenitor property distribution using \texttt{LtsFit} \citep{cappellari_atlas3d_2013} yields a mass--energy relation of $E \propto M^2$ for metal-free \ac{CCSN}e.
		This result agrees well with both observational data and theoretical predictions at solar metallicity \citep[e.g.][]{poznanski_emerging_2013, fang__inferring_2023}. 
		The mass--energy relation appears consistent at zero and solar metallicity. 
		
		Since the abundances observed in EMP stars reflect yields only from successfully exploding stars, we revise the initial mass function (IMF) and explosion energy distribution function (EDF) by introducing a mass-dependent explodability filter $\zeta(M)$. 
		We relax the assumption of independent power-law distributions for mass and energy, and their exponents are constrained by the inferred mass--energy relation. 
		We hypothesize two "islands of explodability", based on two concentrated intervals in the derived mass distribution.
		The first one is between $10-15\,M_\odot$ and the second between $30-40\,M_\odot$. 
		Broadening effects due to model resolution and observational uncertainty are accounted for by convolving the modified distributions with Gaussian kernels and fitting them in cumulative distribution function form. 
		This process yields the best-fit exponents of $\alpha_m=0.54$ and $\alpha_e=0.72$ for IMF and EDF respectively, which is in good agreement with theoretical expectations of $2\alpha_e=\alpha_m+1$ under the derived mass--energy relation. 
		
		In summary, our results demonstrate the utility of the chemical abundance fitting analysis under a Bayesian framework as a robust and independent diagnostic of supernovae physics. 
		This approach complements constraints on supernovae derived from transient observations. 
		EMP stars provides the most readily accessible and reliable observational constraint on Pop III stars, their supernova explosions and early chemical enrichment histories at present. 
		This study underscores the importance of homogeneous analyses of high-resolution spectra and the need for consistent abundance measurements in large samples for future work.
		It also demonstrates the significance of accounting for \ac{NLTE} effects for certain elements, such as Na and Al.
		While the current limitation in sample size, continued advances in the fields of NLTE modeling and nucleosynthesis simulations promise to significantly enhance our understanding of the first generation of stars and the chemical enrichment of the early Universe.
		
		\begin{acknowledgments}
		
		This study is supported by the National Natural Science Foundation of China under grant Nos. 12588202, 12222305 and 12422304, the National Key R\&D Program of China Nos. 2024YFA1611903, 2023YFE0107800 and 2024YFA1611904, and the Strategic Priority Research Program of Chinese Academy of Sciences, grant No. XDB1160301.
		We thank the anonymous referee for their constructive comments and helpful suggestions, which have improved the clarity and quality of this manuscript.
		
		This work has made use of data from the European Space Agency (ESA) mission {\it Gaia} (\url{https://www.cosmos.esa.int/gaia}), processed by the {\it Gaia} Data Processing and Analysis Consortium (DPAC, \url{https://www.cosmos.esa.int/web/gaia/dpac/consortium}). Funding for the DPAC has been provided by national institutions, in particular the institutions participating in the {\it Gaia} Multilateral Agreement.
		
		We also thank the referee for their careful reading of the manuscript and for providing constructive comments that helped improve this work.
		\end{acknowledgments}
		
		\software{LtsFit \citep{cappellari_atlas3d_2013}} 
		
		\begin{acronym}
			\acro{ZAMS}{zero-age main sequence}
			\acro{EMP} {extremely metal-poor}
			\acro{CEMP} {carbon-enhanced metal-poor}
			\acro{CCSN}{core-collapse supernova}
			\acro{PISN}{pair instability supernova}
			\acro{VMP} {very metal-poor}
			\acro{IMF} {initial mass function}
			\acro{EDF}{energy distribution function}
			\acro{CDF}{cumulative density function}
			\acro{EW}{equivalent width}
			\acro{CoG}{curve of growth}
			\acro{LTE}{local thermodynamic equilibrium}
			\acro{NLTE}{non-local thermodynamic equilibrium}
			\acro{IQR}{interquartile range}
			\acro{RUWE}{Renormalised Unit Weight Error}
		\end{acronym}

		\appendix
		\renewcommand{\thefigure}{\thesection.\arabic{figure}}
		\renewcommand{\thetable}{\thesection.\arabic{table}}
		\setcounter{figure}{0}
		\setcounter{table}{0}
		
		\section{Influence of Atmospheric Parameters} \label{appx:atmparam_influence}
		
		To compare the measured abundances based on different determinations of atmospheric parameters, we implement the \ac{LTE} curve-of-growth from L22 on the MMP sample. \citet{norris_most_2013} determine atmospheric parameters of this sample of stars with different methods. We only use the atmospheric parameters of both $T_\mathrm{eff}$ and $\log g$ determined by spectrophotometric flux and hydrogen line profiles. The $T_\mathrm{eff}$ calibration from H $\delta$ line index is not used for this comparison. Specifically, this difference in methods leads to average dispersion of $\Delta T_\mathrm{eff}=128\,\mathrm{K}$ and $\Delta\log g=0.2$ respectively. To isolate the influence of other atmospheric parameters, we fix the metallicity $\mathrm{[Fe/H]}$ and microturbulence velocity ($\xi_t$) as measured by \citet{yong_most_2013}, despite that they might inherently depend on the adopted $T_\mathrm{eff}$ and $\log g$ values. We cataloged the measured abundances and the offset in atmospheric parameters in Table~\ref{tab:compare_atm_impact}. The variation in abundance could be easily computed by $\Delta A(X)=A(X)^s-A(X)^h$, where $s$ and $h$ signify the adopted atmospheric parameters are determined by spectrophotometric flux and hydrogen line profiles respectively. 
		
		This analysis reveals a robust moderate positive correlation with $p\ll0.05$ between temperature offsets $\Delta T_\mathrm{eff}$ and abundance variations $\Delta A(X)$ with Pearson correlation coefficients of $r=0.90, 0.59\ \mathrm{and}\ 0.76$ for Na, Mg and Al consistently. For $\log g$, only a slight negative correlation is found in Mg with $r=-0.46$, whereas no statistically significant relation emerge for bot Na and Al. It indicates the importance of the accurate and precise determination of $T_\mathrm{eff}$ for achieving consistency in abundances across different surveys. 
		
		\startlongtable
		\begin{deluxetable*}{lcccccccccccc}
			\setcounter{table}{2}
			\renewcommand\thetable{\arabic{table}}
			\tablecaption{Impact of atmospheric parameters on abundances}
			\tablehead{
				\vspace{-0.25cm}& &\multicolumn{3}{c}{$\mathrm{A(X)}^s$} & &\multicolumn{3}{c}{$\mathrm{A(X)}^h$} & & & \\
				\colhead{Object name} \vspace{-0.25cm} & & & & & & & & & &$\Delta T_\mathrm{eff}^\dagger$ &$\Delta \log g^\dagger$\\
				\cline{3-5} \cline{7-9}
				& &Na &Mg &Al & &Na &Mg &Al & & &
			}
			\startdata
			52972-1213-507  & &--     &--     &--     & &-- &$4.82$ &-      & &--    &--  \\
			53327-2044-515  & &$2.28$ &$4.07$ &$2.23$ & &$2.31$ &$3.87$ &$2.26$ & &$10$  &$-1.0$\\
			53436-1996-093  & &--     &--     &--     & &$2.60$ &$4.34$ &--     & &--    &--  \\
			54142-2667-094  & &--     &$4.81$ &$2.78$ & &--     &$4.85$ &$2.77$ & &$15$  &$0.2$ \\
			BS 16545-089    & &$2.77$ &$4.59$ &--     & &$2.56$ &$4.39$ &--     & &$278$ &$0.2$ \\
			\enddata
			\label{tab:compare_atm_impact}
			\tablenotetext{s}{For this set of results, atmospheric parameters are determined by spectrophotometric flux}
			\tablenotetext{h}{For this set of results, atmospheric parameters are determined by hydrogen line profiles}
			\tablenotetext{\dagger}{$\Delta T_\mathrm{eff}=T_\mathrm{eff}^s-T_\mathrm{eff}^h$, $\Delta \log g=\log g^s-\log g^h$}
			\tablecomments{Table~\ref{tab:compare_atm_impact} is published in its entirety in the machine-readable format. A portion is shown here for guidance regarding its form and content.}
		\end{deluxetable*}

		\section{NLTE correction and comparison}
		\label{appx:nlte_corr}

		\begin{figure*}
			\centering
			\includegraphics[width=1.\linewidth]{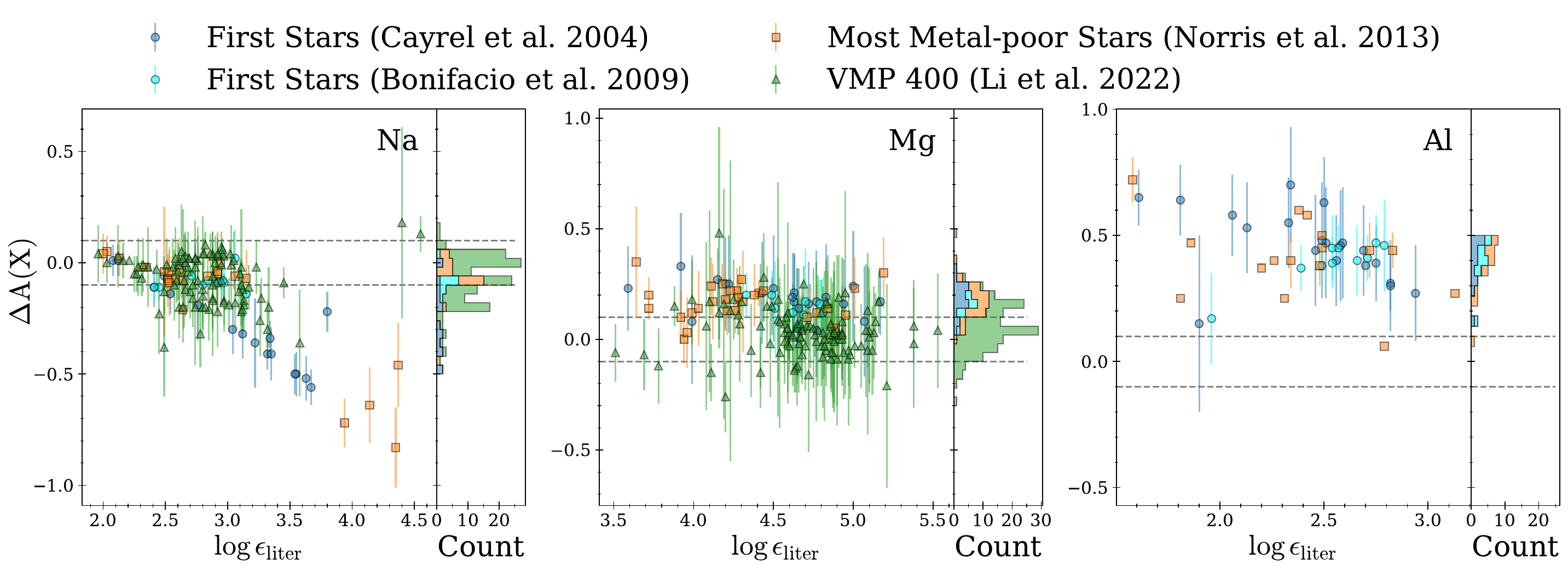}
			\caption{
				\ac{NLTE} correction $\mathrm{A(X)}$ with its corresponding \ac{NLTE} uncertainty as a function of $\log\epsilon$ abundance in the literature. The color coding is the same as Figure~\ref{fig:abudist}.
			}
			\label{fig:nltecorr_nlte_lte}
		\end{figure*}
		
		Post-analysis \ac{NLTE} correction requires both the measurement of \ac{EW} and the determination of atmospheric parameters. We first examine the relationship between \ac{EW} and the NLTE-corrected elemental abundance. For a given set of stellar atmospheric parameters in the L22 grids, the elemental abundance $\log\epsilon$ generally exhibits an approximate linear relationship with $\log\ac{EW}$. Accordingly, the abundance--\ac{EW} relation $\log\epsilon(\log\ac{EW})$ could be easily obtained through linear interpolation. Second, to simplify the dependence of atmospheric parameters, we approximate $\log\epsilon(\log\ac{EW})$ as a linear functional in a four-dimensional space defined by $\left(T_\mathrm{eff}, \log g, \mathrm{[Fe/H]}, \xi_t\right)$. For a given set of stellar parameter offsets $\Delta\left(T_\mathrm{eff}, \log g, \mathrm{[Fe/H]}, \xi_t\right)$ from the nearest grid note, denoted by subscript $0$, the \ac{NLTE}-corrected abundance  could be expressed as:
		\begin{eqnarray}
			\log\epsilon(\log\ac{EW}; T_\mathrm{eff}, \log g, \mathrm{[Fe/H]}, \xi_t) = \log\epsilon(\log\ac{EW}; T_{\mathrm{eff}, 0}, \log g_0, \mathrm{[Fe/H]_0}, \xi_{t, 0}) + \nonumber \\ 
			\Delta T_\mathrm{eff} \left. \frac{\mathrm{d}\log\epsilon}{\mathrm{d}T_\mathrm{eff}} \right|_0 + \Delta \log g \left. \frac{\mathrm{d}\log\epsilon}{\mathrm{d}\log g} \right|_0+ \Delta \mathrm{[Fe/H]} \left. \frac{\mathrm{d}\log\epsilon}{\mathrm{d}\mathrm{[Fe/H]}} \right|_0 + \Delta \xi_{t} \left. \frac{\mathrm{d}\log\epsilon}{\mathrm{d}\xi_{t}} \right|_0. 
		\end{eqnarray}
		
		The uncertainty for the \ac{NLTE} abundance in this work only accounts for two components \citep{barklem_hamburgeso_2005}: the propagated error from the uncertainties in atmospheric parameters of $T_\mathrm{eff}$, $\log g$ and $\xi_t$, and the observational error across different spectral lines. By shifting the atmospheric parameters by $1\sigma$ uncertainty of $\sigma_{T_\mathrm{eff}}=100\,\mathrm{K}$, $\sigma_{\log g}=0.2$ and $\sigma_{\xi_t}=0.2\,\mathrm{km\,s^{-1}}$ respectively, we could estimate the measurement error caused by each parameter. If the covariance of parameters is neglected, the first part of error could be given by the quadratic addition of the errors of atmospheric parameters. 
		As for the second term of observational uncertainty, it could be estimated with an unbiased standard deviation:
		\begin{equation}
			\sigma_{N_\mathrm{lines}}=\sqrt{ \frac{\sum_{i=1}^{N_\mathrm{lines}}(\log\epsilon_i-\overline{\log\epsilon})} {N_\mathrm{lines}-1} },
		\end{equation}
		where $N_\mathrm{lines}$ denotes the number of spectral lines used in the abundance determination. 
		However, due to the presence of line-to-line variations, elemental abundances derived from different absorption lines may exhibit noticeable discrepancies. To mitigate the influence of outliers, we exclude absorption lines whose abundances deviate beyond 1.5 times the \ac{IQR} from the median value when the number of absorption lines $N_\mathrm{lines}>3$. In cases where $N_\mathrm{lines} = 3$, an absorption line is flagged as an outlier if its abundance differs from the remaining two by more than twice the abundance difference between the latter two lines. For cases with only $N_\mathrm{lines} = 2$, where it is not possible to unambiguously identify the outlier, the abundance is considered unreliable and discarded if the unbiased standard deviation exceeds $0.3\,\mathrm{dex}$.
		
		\begin{deluxetable}{lccccccccccccccccccccc}
			\setcounter{table}{3}
			\renewcommand\thetable{\arabic{table}}
			\tablecolumns{22}
			\tablecaption{\ac{NLTE} Correction for the Selected Stars\label{tab:nltecorr}}
			\tablehead{
				\vspace{-.25cm}& &\multicolumn{3}{c}{$\mathrm{A(X)^n}$} & &\multicolumn{3}{c}{$\mathrm{A(X)^l}$}  & &\multicolumn{3}{c}{$\Delta \mathrm{A(X)}$} \\
				\colhead{Object name}\vspace{-.25cm} \\
				\cline{3-5} \cline{7-9} \cline{11-13}
				& &Na &Mg &Al & &Na &Mg &Al & &Na &Mg &Al
			}
			\startdata
			BD-18 5550   &  & 2.91 & 5.02 & 3.13 &  & 3.32 & 4.83 & 2.82 &  & $-0.41$ & $+0.19$ & $+0.31$ \\
			BS 16477-003 &  & 2.74 & 4.73 & 2.98 &  & 3.04 & 4.5  & 2.51 &  & $-0.30$  & $+0.23$ & $+0.47$ \\
			CS 22172-002 &  & 2.13 & 4.25 & 2.45 &  & 2.12 & 3.92 & 1.81 &  & $+0.01$  & $+0.33$ & $+0.64$ \\
			CS 22186-025 &  & 3.11 & 5.1  & 3.13 &  & 3.67 & 4.94 & 2.69 &  & $-0.56$ & $+0.16$ & $+0.44$
			\enddata
			\tablenotetext{n}{\ac{NLTE}-corrected abundance for element $\mathrm{X}$}
			\tablenotetext{l}{Abundance in literature for element $\mathrm{X}$}
			\tablecomments{Table~\ref{tab:nltecorr} is published in its entirety in the machine-readable format. A portion is shown here for guidance regarding its form and content.}
		\end{deluxetable}
		
		\startlongtable
		\begin{splitdeluxetable*}{lcccccccccccBcccccccc}
			\setcounter{table}{4}
			\renewcommand\thetable{\arabic{table}}
			\tablecolumns{20}
			\tablecaption{$\log\ac{EW}$s of the Used Abosption Lines for the Selected Stars\label{tab:equivwidth}}
			\tablehead{
				\vspace{-.25cm}& &\multicolumn{2}{c}{$\log\ac{EW}_\mathrm{Na}$~[m\AA]} & &\multicolumn{7}{c}{$\log\ac{EW}_\mathrm{Mg}$~[m\AA]} &\multicolumn{5}{c}{$\log\ac{EW}_\mathrm{Mg}$~[m\AA]}  & &\multicolumn{2}{c}{$\log\ac{EW}_\mathrm{Al}$~[m\AA]} \\
				\colhead{Object name}\vspace{-.25cm} \\
				\cline{3-4} \cline{6-17} \cline{19-20}
				&  & 5890 & 5896 &  & 3829 & 3832 & 3838 & 4058 & 4167 & 4352 & 4571 & 4703 & 5173 & 5184 & 5528 & 5711 &  & 3944 & 3962 
			}
			\startdata
			BD-18 5550   &  & 2.14 & 2.06 &  & 2.18 & 2.24 & 2.28 &-- & 1.57 & 1.68 & 1.57 &-- & 2.21 & 2.25 & 1.71 &-- &  & 2.00  & 2.06 \\
			BS 16477-003 &  & 2.07 & 1.95 &  & 2.10 & 2.14 & 2.16 &-- & 1.33 & 1.48 & 1.28 &-- & 2.11 & 2.14 & 1.45 &-- &  & 1.91 & 1.96 \\
			CS 22172-002 &  & 1.82 & 1.65 &  & 2.02 & 2.08 & 2.12 &-- & 0.91 & 0.94 &--  &-- & 2.04 & 2.09 & 1.02 &-- &  & 1.70 & 1.81 \\
			CS 22186-025 &  & 2.20 & 2.10 &  & 2.19 & 2.25 & 2.28 &-- &-- & 1.62 & 1.54 &-- & 2.23 & 2.25 & 1.68 &-- &  & 1.95 & 2.03\\
			\enddata
			\tablecomments{Table~\ref{tab:equivwidth} is published in its entirety in the machine-readable format. The wavelengths of absoption lines are in units of Angstroms.}
		\end{splitdeluxetable*}
		
		We present the adopted \ac{NLTE} corrections in Figure~\ref{fig:nltecorr_nlte_lte}. Furthermore, the corresponding data for the selected 40 stars are summarized in Table~\ref{tab:nltecorr}, and the adopted equivalent widths in the literature are cataloged in Table~\ref{tab:equivwidth}.
		The \ac{NLTE} correction is computed as $\Delta \mathrm{A(X)} = \mathrm{A(X)^n} - \mathrm{A(X)^l}$, where superscript $n$ and $l$ signify \ac{NLTE}-corrected and literature abundance respectively. It is important to note that the element abundances in the literature are mostly measured based on \ac{LTE} assumptions, with the exception of Na abundances in \citet{li_four-hundred_2022}, which have been corrected for \ac{NLTE} effects using the earlier grids of \citet{lind_non-lte_2011}. 
		
		\bibliography{BayesFit}
		\bibliographystyle{aasjournal}
		
	\end{CJK*}
\end{document}